\documentclass[twocolumn]{aa}

\usepackage{graphicx}
\usepackage{amsmath}
\usepackage{amssymb}
\usepackage{bm}
\usepackage{txfonts} 
\usepackage{mathtools}
\usepackage{xcolor}
\usepackage{hyperref}
\usepackage{fontawesome}
\usepackage{ulem}

\newcommand{\mat}[1]{\bm{\mathrm{#1}}}                                          
\renewcommand{\vec}[1]{\bm{#1}}                                                 

\newcommand*{\Euclid}{\textit{Euclid}\xspace}
\newcommand{\ep}{{e^{\textrm{p}}}}                                                
\newcommand{\Tp}{{T^{\textrm{p}}}}

\DeclareMathOperator*{\argmin}{arg\,min}

\newcommand{\bin}[2]{\Delta_\vartheta(\vartheta_{#1 #2})}

\begin{document}

   \title{Galaxy-Point Spread Function correlations as a probe of weak-lensing systematics with UNIONS data}


   \author{%
    Sacha Guerrini
    \inst{1}
    \fnmsep\thanks{\email{sacha.guerrini@cea.fr}}
    \and
    Martin Kilbinger
    \inst{2}
    \fnmsep\thanks{\email{martin.kilbinger@cea.fr}}
    \and
    Hubert Leterme
    \inst{3, 2}
    \and
    Axel Guinot
    \inst{4}
    \and
    Jingwei Wang
    \inst{2,5,6}
    \and
    Fabian Hervas Peters
    \inst{2}
    \and
    Hendrik Hildebrandt
    \inst{11}
    \and
    Michael J.~ Hudson
    \inst{7, 8, 9}
    \and 
    Alan McConnachie
    \inst{10}
    }

   \institute{%
   Université Paris Cité, Université Paris-Saclay, CEA, CNRS, AIM, F-91191, Gif-sur-Yvette, France
   \and
   Université Paris-Saclay, Université Paris Cité, CEA, CNRS, AIM, 91191, Gif-sur-Yvette, France
   \and
   Université Caen Normandie, ENSICAEN, CNRS, Normandie Univ, GREYC UMR 6072, F-14000 Caen, France
   \and
   Department of Physics, McWilliams Center for Cosmology, Carnegie Mellon University, Pittsburgh, PA 15213, USA
   \and
   Sorbonne Université, CNRS, UMR 7095, Institut d’Astrophysique de Paris (IAP), 98 bis boulevard Arago, 75014 Paris, France
   \and
   École Nationale Supérieure des Mines de Paris, 60 boulevard Saint-Michel, 75272 Paris Cedex 06, France
   \and
   Department of Physics and Astronomy, University of Waterloo, Waterloo, ON, N2L 3G1, Canada
   \and
   Waterloo Centre for Astrophysics, Waterloo, ON, N2L 3G1, Canada
   \and
   Perimeter Institute for Theoretical Physics, 31 Caroline St. N., Waterloo, ON, N2L 2Y5, Canada
   \and
   NRC Herzberg Astronomy and Astrophysics Research Centre, 5071 West Saanich Road, Victoria, B.C., Canada, V9E 2E7
   \and
   Ruhr University Bochum, Faculty of Physics and Astronomy, Astronomical Institute (AIRUB), German Centre for Cosmological
Lensing, 44780 Bochum, Germany
    }

   \date{Received XXX; accepted YYY}

 
  \abstract
   {
    Weak gravitational lensing requires precise measurements of galaxy shapes and therefore an accurate knowledge of the PSF model. The latter can be a source of systematics that affect the shear two-point correlation function. A key stake of weak lensing analysis is to forecast the systematics due to the PSF.
   }
   {
   Correlation functions of galaxies and the PSF, the so-called $\rho$- and $\tau$-statistics, are used to evaluate the level of systematics coming from the PSF model and PSF corrections, and contributing to the two-point correlation function used to perform cosmological inference. Our goal is to introduce a fast and simple method to estimate this level of systematics and assess its agreement with state-of-the-art approaches.
   }
   {
   We introduce a new way to estimate the covariance matrix of the $\tau$-statistics using analytical expressions. The covariance allows us to estimate parameters directly related to the level of systematics associated with the PSF and provides us with a tool to validate the PSF model used in a weak-lensing analysis. We apply those methods to data from the Ultraviolet Near-Infrared Optical Northern Survey (UNIONS).
   }
   {
   We show that the semi-analytical covariance yields comparable results than using covariances obtained from simulations or jackknife resampling. It requires less computation time and is therefore well suited for rapid comparison of the systematic level obtained from different catalogs. We also show how one can break degeneracies between parameters with a redefinition of the $\tau$-statistics.
   }
   {
   The methods developed in this work will be useful tools in the analysis of current weak-lensing data but also of Stage IV surveys such as \textit{Euclid}, LSST or \textit{Roman}. They provide fast and accurate diagnostics on PSF systematics that are crucial to understand in the context of cosmic shear studies.
   }

   \keywords{
    cosmology-cosmic shear-
    PSF diagnostics-$\tau$-statistics
    }

   \maketitle
%

\section{Introduction}

Weak gravitational lensing is a powerful probe to study the distribution of mass in the Universe and to constrain cosmological models. Light emitted by distant galaxies is warped by the gravitational field due to over- and underdensities along the line of sight. This effect is typically of the order of a few percent but can be observed as coherent shape distortions of source galaxies called ``cosmic shear'' \citep[for reviews see, e.g.,][]{BS01,K15,2017arXiv171003235M}. In the past decade, Stage-III photometric surveys such as the Dark Energy Survey \citep[DES;][]{descollaborationDarkEnergySurvey2022}, the Kilo-Degree Survey \citep[KiDS;][]{asgariKiDS1000CosmologyCosmic2021a}, the Hyper Suprime-Cam survey \citep[HSC;][]{dalalHyperSuprimeCamYear2023, liHyperSuprimeCamYear2023} provided constraints on cosmological parameters from cosmic shear. The Ultraviolet Near-Infrared Optical Northern Survey \citep[UNIONS;][]{UNIONS_Guinot_SP} cosmological analysis is currently in progress and will perform comparable analyses. Forthcoming Stage-IV surveys, such as \Euclid \citep{2011arXiv1110.3193L, euclidcollaborationEuclidOverviewEuclid2024a}, the Vera Rubin Observatory Legacy Survey of Space and Time \citep[LSST;][]{ivezicLSSTScienceDrivers2019} or the \textit{Nancy Grace Roman} Space Telescope \citep{akesonWideFieldInfrared2019}, will measure cosmic shear with reduced statistical uncertainties thanks to a large area coverage and great depth resulting in a high number density of observed galaxies. Obtaining reliable cosmological results from cosmic shear therefore imposes stringent requirements on controlling and modelling systematic biases and uncertainties that affect cosmic shear measurements.

In practice, for cosmic shear one must precisely measure the shapes of galaxies observed in the field of view. However, optical systems can distort the true images of galaxies according to its Point Spread Function (PSF) \citep{mandelbaumWeakLensingPrecision2018}. The PSF describes the image response to the light of a point source, after passing through atmospheric turbulence and the telescope optics. The observed images correspond to the convolution of the PSF with the images of all observed objects. The PSF can thus modify the shape of galaxies in a systematic way and induce biases and uncertainties in the measurement of galaxy shapes. This effect can be modelled via multiplicative and additive biases that depend on the size and anisotropies of the PSF \citep{kaiserMethodWeakLensing1995, Heymans06}. An error in the size of the PSF results in a multiplicative bias due to an incorrect estimation of how the PSF has rounded the object. Likewise, an error in the PSF shape can give rise to both multiplicative and additive biases. The former can be induced by a coherent alignment or misalignment of the PSF with galaxy shapes even with a perfect PSF model and the latter occurs due to modelling errors. The effect of the shape and size mismodelling can be tested statistically using the so-called $\rho$-statistics
\citep{2008A&A...484...67P, 2016MNRAS.460.2245J} and the $\tau$-statistics \citep{10.1093/mnras/stab918}. The $\rho$-statistics are the correlations between the PSF ellipticity and its shape and size residuals, whereas the $\tau$-statistics cross-correlate those PSF fields with galaxy shapes. The $\tau$-statistics provide a null test to identify additive or multiplicative biases but can also be used to estimate the level of systematic uncertainty that propagate to the two-point correlation function, allowing to forward model the systematic error on the two-point correlation function (2PCF) for cosmological inference.

This estimation of the systematic level requires solving a linear problem which expresses the $\tau$-statistics as functions of the $\rho$-statistics and parameters of interest. Estimating those parameters therefore requires a reliable estimate of the covariance matrix of the $\tau$-statistics. Previous work used either simulations \citep{zhangGeneralFrameworkRemoving2023} or jackknife resampling \citep{10.1093/mnras/stab918} to perform this estimation. The use of simulations, on the one hand, does not easily allow us to capture the whole covariance as we only simulate galaxies (and not stars) to estimate the covariance. On the other hand, jackknife resampling is based on the prior that our measurement of the $\tau$-statistics is sampled from the correct distribution. It also requires to build patches of the sky which can be cumbersome in the presence of masks. Finally, the covariance obtained with jackknife resampling tends to be noisier than the one obtained from simulations. It will be useful for future surveys to be able to perform fast PSF diagnostics i.e.~a quick estimation of the $\tau$ covariance matrix to compare galaxy catalogs obtained under different modelling choices. 

In this paper, we introduce a semi-analytical method to compute the covariance matrix of the $\tau$-statistics and therefore estimate the level of systematic error on the two-point correlation function. We also show how least-square minimization provides a fast estimation of systematics parameters and their uncertainties. In addition, we discuss how one can break degeneracies in the parameters defining the PSF error model. This paper is organised as follows. In Sect.~\ref{Section 2: Method} we review the $\rho$- and $\tau$- statistics and introduce our semi-analytical estimate of the covariance matrix for the $\tau$-statistics. We also present our sampling methods of the parameters. In Sect.~\ref{seq: data and simulations}, we introduce the catalogs on which we test this method, and the simulations used to estimate the covariance matrix. We demonstrate in Sect.~\ref{seq: results} that our semi-analytical covariance is in agreement with state-of-the-art jackknife resampling and simulation-based estimates. We discuss the performance and features of our PSF diagnostics estimation in Sect.~\ref{seq:discussion}. In addition, we present a reformulation of PSF systematics that reduce degeneracies in the parameters of the model. The paper concludes in Sect.~\ref{seq: Conclusions}.

\section{Method}\label{Section 2: Method}

In this section, we first introduce our general notation for spin-$2$ fields, their correlation functions and estimators (see Sect.~\ref{seq:estimators}).
We then introduce shear and PSF (residual) fields and their relation to observable quantities, before reviewing the $\rho$- and $\tau$-statistics (see Sects.~\ref{seq:estimators}-\ref{seq:rho_stats}). We discuss the inference problem of the PSF contamination parameters, and introduce our least-squares approach to solve it (see Sect.~\ref{seq:tau_stats}). Lastly, we develop the semi-analytical covariance for the $\tau$-statistics (see Sect.~\ref{seq: analytical covariance}).

\subsection{Second-order correlations and estimators for spin-$2$ fields}\label{seq:estimators}

For two statistically isotropic and homogeneous random fields $a$ and $b$, their joint two-point correlation between two angular positions $\vec \theta$ and $\vec \theta + \vec \vartheta$, $\left\langle a(\vec \theta) b(\vec \theta + \vec \vartheta) \right\rangle$, only 
depends on the modulus of the vector between the two positions, $\vartheta$. Under the ergodic principle, the ensemble average can be replaced with a spatial average, and we write
\begin{equation}
    \left\langle a b \right\rangle(\vartheta)
    = \left\langle a(\vec \theta) b(\vec \theta + \vec \vartheta) \right\rangle_{\vec \theta} .
\end{equation}
Shear and ellipticity are spin-$2$ fields, which can be written in complex notation as
$a = a_1 + \textrm{i} a_2 = |a| \exp(2 \textrm{i} \varphi)$. Here, the real (resp.~imaginary) part of the field, $a_1$ (resp.~$a_2$), quantifies the amplitude of the ellipticity of the object oriented at a $0$ (resp.~$45$) degree angle with respect to the $x$-axis in a local Cartesian coordinate system. The phase $\varphi$ denotes the orientation of the ellipse. 

For a pair of galaxies with connecting vector $\vec \vartheta$ and polar angle $\phi$, the shear of both galaxies is conveniently rotated into the coordinate system where $\vec \vartheta$ is the first coordinate direction. This defines the tangential and cross-components of the shear, $a_\textrm{t}$ and $a_\times$, respectively, as
\begin{align}
    a_\textrm{t} = \Re \left( a \textrm{e}^{- 2 \textrm{i} \phi} \right); \quad
    a_\times = \Im \left( a \textrm{e}^{- 2 \textrm{i} \phi} \right).
    \label{eq:a_tx}
\end{align}
The two standard parity-invariant correlation functions \citep{2002A&A...389..729S} for the spin-$2$ fields $a$ and $b$
are defined as
\begin{align}\label{eq: xi_plus_minus}
    \xi_+^{ab}(\vartheta)
    & = 
    \left\langle a_\textrm{t} b_\textrm{t} \right\rangle(\vartheta)
    + \left\langle a_\times b_\times \right\rangle(\vartheta)
    \nonumber \\
    &=
    \left\langle a_1 b_1 \right\rangle(\vartheta)
    + \left\langle a_2 b_2 \right\rangle(\vartheta)
    \nonumber \\
    &= \Re \left[ \langle ab^* \rangle (\vartheta) \right]
    ;
    \nonumber \\
    \xi_-^{ab}(\vartheta)
    & = 
    \left\langle a_\textrm{t} b_\textrm{t} \right\rangle(\vartheta)
    - \left\langle a_\times b_\times \right\rangle(\vartheta)
    \nonumber \\
    &=
    \left\langle
    \left[
    \left( a_1 b_1 \right)(\vartheta)
    - \left( a_2 b_2 \right)(\vartheta)
    \right] \cos 4 \phi
    \right\rangle
    \nonumber \\
    & \quad +
    \left\langle
    \left[
    \left( a_1 b_2 \right)(\vartheta)
    + \left( a_2 b_1 \right)(\vartheta)
    \right] \sin 4 \phi
    \right\rangle
    \nonumber \\
    &= \Re \left\langle \left( ab \right) (\vartheta) e^{- 4 \textrm{i} \phi} \right\rangle .
\end{align}
Parity-violating correlation functions are reproduced in App.~\ref{sec:terms_der}.
Following \cite{SvWKM02}, 
estimators of the correlation functions \eqref{eq: xi_plus_minus} are
\begin{align} 
\hat \xi_+^{ab}(\vartheta)
    = &
    \frac 1 {N_\textrm{p}^{ab}(\vartheta)}
      \sum_{ij} w^a_i w^b_j
      \bin{i}{j} 
        \left( a_{i1} b_{j1} + a_{i2} b_{j2} \right)
    \nonumber\\
    = &
     \frac 1 {N^{ab}_\textrm{p}(\vartheta)}
      \sum_{ij} w^a_i w^b_j
      \bin{i}{j} 
      \sum_{\alpha=1}^2
        a_{i \alpha} b_{j \alpha}
      ;
      \nonumber
      \\
\hat \xi_-^{ab}(\vartheta)
     = &
    \frac 1
    {N^{ab}_\textrm{p}(\vartheta)}
      \sum_{ij} w^a_i w^b_j
      \bin{i}{j} 
      \nonumber \\
      & \times \left[
        \left( a_{i1} b_{j1} - a_{i2} b_{j2} \right) \cos 4 \phi_{ij}
        + 
        \left( a_{i1} b_{j2} + a_{i2} b_{j1} \right) \sin 4 \phi_{ij}
      \right]
    .
    \label{eq:hat_xi_pm}
\end{align}
Here, $N_\textrm{p}^{ab}(\vartheta)$ is the number of pairs of objects sampling the fields $a$ and $b$ in the angular bin around $\vartheta$, given as
\begin{align}
    N_\textrm{p}^{ab} = \sum_{ij} w^a_i w^b_j \bin{i}{j}.
\end{align}
The sums are carried out over all objects $i$ sampling field $a$, and all objects $j$ sampling field $b$, whose pair-wise distance $\vartheta_{ij}$ is within the angular bin around $\vartheta$. The bin is given by $\bin{i}{j} = 1_{[\vartheta - \Delta\vartheta/2; \vartheta+\Delta\vartheta/2]}(\vartheta_{ij})$, where the indicator function $1_S(x)$ is $1$ if $x \in S$ and $0$ otherwise.

The ellipticity or shear of each object $a_i$ (resp.~$b_j$) has an associated weight $w^a_i$ (resp.~$w^b_j$). Those weights account for variable signal-to-noise ratios for the different objects that introduce uncertainty in the shape measurement \citep[see, e.g.,][]{10.1093/mnras/stab918}.
 
Hereafter, we reproduce useful relations adapted from \cite{SvWKM02} for the correlation of two spin-2 fields $a$ and $b$. Their derivation with the inclusion of parity-violating correlations is detailed in App.~\ref{sec:terms_der}.
\begin{align}
    \langle a_{i1} b_{j1} \rangle & =
        \frac{1}{2} \left[
            \xi_+^{ab}(\vartheta_{ij})
            + \xi_-^{ab}(\vartheta_{ij}) \cos 4\phi_{ij}
            - \xi_\times^{ab}(\vartheta_{ij}) \sin 4 \phi_{ij}
        \right] ;
   \nonumber \\
    \langle a_{i2} b_{j2} \rangle & =
        \frac{1}{2} \left[
            \xi_+^{ab}(\vartheta_{ij})
            - \xi_-^{ab}(\vartheta_{ij}) \cos 4\phi_{ij}
            + \xi_\times^{ab}(\vartheta_{ij}) \sin 4 \phi_{ij}
            \right];
    \nonumber \\
    \langle a_{i1} b_{j2} \rangle & =
        \frac{1}{2} \left[
            \xi_-^{ab}(\vartheta_{ij}) \sin 4\phi_{ij}
            + \xi_\times^{ab}(\vartheta_{ij}) \cos 4 \phi_{ij}
            - \xi_\ast^{ab}(\vartheta_{ij})
        \right];
    \nonumber \\
     \langle a_{i2} b_{j1} \rangle & =
        \frac{1}{2} \left[
            \xi_-^{ab}(\vartheta_{ij}) \sin 4\phi_{ij}
            + \xi_\times^{ab}(\vartheta_{ij}) \cos 4 \phi_{ij}
            + \xi_\ast^{ab}(\vartheta_{ij})
        \right] .
    \label{eq:terms_ab}
\end{align}
Under the assumption that $\xi_\times$ (see Eq.~\eqref{eq: xi_star_cross}) vanishes and with $a = b$, the first two equations reduce to Eq.~(22) in \cite{SvWKM02}. With additionally $\xi_\ast = 0$, the sum of our third and fourth equations equals twice the mixed expression from their Eq.~(22). In what follows, we will keep those assumptions, corresponding to parity conservation of the involved fields.

\subsection{Shear and PSF (residual) ellipticity}\label{seq:shear_psf_ell}

In general, shear $\gamma$ is estimated by the observed shape of background galaxies, quantified by an ellipticity measurement $e^\textrm{obs}$. This is a noisy estimate, with the intrinsic (source) galaxy ellipticity $e^\textrm{s}$ as a stochastic element. We assume an additive systematic contribution, $e^\textrm{sys}$ stemming from the PSF, and write
\begin{equation}
    e \equiv e^\textrm{obs} = e^\textrm{s} + e^\textrm{PSF, sys} + \gamma.
    \label{eq:eobs}
\end{equation}
Here, we assume that the observed ellipticity has been calibrated for multiplicative biases.
While we expect $\langle e^{\rm s} \rangle = 0$ in the absence of intrinsic alignment, a non-zero $\langle e^{\rm PSF, sys} \rangle$ would be a smoking gun for a systematic PSF contribution to the measured shapes of galaxies. A commonly used linear model for the PSF contribution is
\begin{align}
    e^\textrm{PSF, sys} = \alpha \ep + \beta \, \delta \ep + \eta \, \delta \Tp,
    \label{eq:e_sys}
\end{align}
where $\ep$ is the ellipticity of the PSF, $\delta \ep = e^* -\ep$ denotes the ellipticity residual (measured at star positions), and $\delta \Tp = e^* (T^* - \Tp)/T^*$ is the size residual, also defined at star positions. The size term is multiplied by the star ellipticity $e^*$ to define a spin-$2$ field, consistent with \cite{2008A&A...484...67P}.
Other models can and have been used as well \citep[e.g.][]{giblinKiDS1000CatalogueWeak2021a, zhangGeneralFrameworkRemoving2023,ZK24}.
The first term in Eq.~\eqref{eq:e_sys} corresponds to a spurious dependence of the measured shape on the PSF, the so-called PSF leakage. Such a leakage might be introduced by an imperfect deconvolution of the observed galaxy image by the PSF. The second and third terms in the above equation quantify PSF modelling and interpolation errors.

The PSF and its residuals are estimated from measurements of the ellipticity of stars and the PSF model. The PSF ellipticity and size are exact and deterministic once the PSF model is fixed. The ellipticity and size of stars are measured with a small amount of stochasticity since stars are point sources, and are typically selected at a high signal-to-noise ratio. In addition, there is pixel noise, unresolved binaries, cosmic rays, saturation effects, etc. Those effects introduce a stochastic element in measuring star ellipticity.

Likewise, the atmosphere for ground-based observations is stochastic but this stochasticity is part of the PSF at a given observation epoch, sky and focal plane position; the atmosphere is part of the ``signal'' and not the noise.

The estimate for the shot noise for galaxies, $\sum_i |e_i|^2/N$ is a very good approximation of $\langle |e^\textrm{s}|^2 \rangle$, since $\langle |\gamma|^2 \rangle$ is much smaller and can fairly safely be neglected.
In contrast, for the reasons stated above, star and PSF parameters are dominated by the signal over the noise. The estimation of shot noise is therefore not straight-forward.
In addition, the size residual is a combination of star ellipticity and size, which might be correlated.

As we will describe in detail in Sect.~\ref{seq:derivation_covariance}, the inclusion of the estimated shot noise for star and PSF parameters leads to an incorrect semi-analytical covariance matrix of the galaxy - PSF cross-correlation functions.
For this reason, unlike in Eq.~\eqref{eq:eobs} where the intrinsic shape of galaxies $e^s$ is taken into account, we do not include a stochastic ellipticity $b^s$ in the relation between observed and true PSF quantities. Instead, the shot noise of the galaxy - PSF cross-correlation functions will be measured from the data and added independently to the semi-analytical covariance matrix.

Here, we write the relation between observed and true star and PSF parameters as
\begin{equation}
    b^\textrm{obs} = b \quad \mbox{for}
    \quad b \in \{ \ep, \delta \ep, \delta \Tp \}.
    \label{eq:bobs}
\end{equation}

\subsection{$\rho$-statistics}\label{seq:rho_stats}

The $\rho$-statistics are cross-correlation functions of the PSF-related fields appearing in the systematic error contribution to Eq.~\eqref{eq:e_sys}. They are defined as follows \citep[see][]{2010MNRAS.404..350R,2016MNRAS.460.2245J,10.1093/mnras/stab918}:
\begin{align}\label{eq:rho_stats}
    \rho_0(\vartheta) & = \langle \ep \ep \rangle(\vartheta); &
    \rho_1(\vartheta) & = \langle \delta \ep \delta \ep \rangle(\vartheta); \nonumber \\
    \rho_2(\vartheta) & = \langle \ep \delta \ep \rangle(\vartheta); &
    \rho_3(\vartheta) & = \langle \delta \Tp \delta \Tp \rangle(\vartheta); \nonumber \\
    \rho_4(\vartheta) & = \langle \delta \ep \delta \Tp \rangle(\vartheta); &
    \rho_5(\vartheta) & = \langle \ep \delta \Tp \rangle(\vartheta).
\end{align}
We abused the notation $\langle . \rangle$ for simplicity. In practice, each $\rho$-statistics has $+$, $-$, $*$ and $\times$ component obtained using Eqs.~\eqref{eq: xi_plus_minus} and \eqref{eq: xi_star_cross} with the corresponding field $a$ and $b$ (e.g. $\rho_{0, \pm} = \xi_\pm^{\rm \ep \ep}$). Those statistics quantify contributions to additive bias of the shear two-point correlation function.

Under the linear model \eqref{eq:e_sys}, we cannot quantify the systematic error with the $\rho$-statistics alone as it depends on the value of the parameters $\alpha$, $\beta$ and $\eta$. Hence, we need to compute cross-correlations between galaxies and the PSF fields to estimate those parameters.

\subsection{$\tau$-statistics}\label{seq:tau_stats}

The $\tau$-statisics \citep{hamanaCosmologicalConstraintsCosmic2020,2021A&A...645A.105G,10.1093/mnras/stab918,zhangGeneralFrameworkRemoving2023} are obtained by cross-correlating the galaxy ellipticities with the different PSF terms appearing in Eq.~\eqref{eq:e_sys}:
\begin{equation}
    \tau_0(\vartheta) = \langle e \, \ep \rangle(\vartheta); \quad
    \tau_2(\vartheta) = \langle e \, \delta \ep \rangle(\vartheta); \quad
    \tau_5(\vartheta) = \langle e \, \delta \Tp \rangle(\vartheta),
    \label{eq:tau}
\end{equation}
where we use the same abuse of notation as above. The galaxy ellipticities $e$ appearing in the definition of the $\tau$-statistics have been calibrated (see Sect.~\ref{seq: data and simulations} for details on the calibration) and mean-substracted to remove a non-zero $\langle \gamma \rangle$ due to cosmic variance. Interestingly, those correlators allow to fit a model to estimate the value of $\alpha$, $\beta$ and $\eta$ using the following linear equations for a given scale $\vartheta$:
\begin{align}\label{eq:linear_system}
  \tau_0(\vartheta) = & \alpha \, \rho_0(\vartheta)
  + \beta \, \rho_2(\vartheta)
    + \eta \, \rho_5(\vartheta);
\nonumber\\
  \tau_2(\vartheta) = & \alpha \, \rho_2(\vartheta)
  + \beta \, \rho_1(\vartheta)
   + \eta \, \rho_4(\vartheta);
\\
 \tau_5(\vartheta) = & \alpha \, \rho_5(\vartheta)
 + \beta \, \rho_4(\vartheta)
  + \eta \, \rho_3(\vartheta). \nonumber
\end{align}
One can choose to invert the system of equations at each scale, or to consider scale-independent parameters $\alpha$, $\beta$ and $\eta$. For simplicity and because the $\rho$- and $\tau$-statistics are noisy, we assume in this work that these parameters are scale-independent. The linear equations per scale \eqref{eq:linear_system} can be concatenated into a single matrix equation:
\begin{equation}
    \left(
        \begin{array}{l}
            \tau_{0,1} \\ \tau_{2,1} \\ \tau_{5, 1} \\ \vdots \\
            \tau_{0, n} \\ \tau_{2, n} \\ \tau_{5, n}
        \end{array}
    \right)
    =
    \left(
      \begin{array}{llllll}
        \rho_{0, 1} & \rho_{2, 1} & \rho_{5, 1} \\
        \rho_{2, 1} & \rho_{1, 1} & \rho_{4, 1} \\
        \rho_{5, 1} & \rho_{4, 1} & \rho_{3, 1} \\
               & \ddots &        \\
        \rho_{0, n} & \rho_{2, n} & \rho_{5, n} \\
        \rho_{2, n} & \rho_{1, n} & \rho_{4, n} \\
        \rho_{5, n} & \rho_{4, n} & \rho_{3, n} \\
      \end{array}
    \right)
    \left(
      \begin{array}{l}
        \alpha \\ \beta \\ \eta \\
      \end{array}
    \right),
    \label{eq:tau_matrix}
\end{equation}
where $\rho_{i,j}$ denotes the value of the $i$-th $\rho$-statistics at the $j$-th angular bin and similarly $\tau_{i,j}$ is the value of the $i$-th $\tau$-statistics at the $j$-th angular bin. Equation \eqref{eq:tau_matrix} writes in short:
\begin{equation}\label{eq:linear_system_matrix_form}
    \vec \tau = \mat R \vec \Omega + \mat \Sigma.
\end{equation}
where $\vec \Omega = (\alpha, \beta, \eta)^\textrm{T}$ and $\mat \Sigma$ is a noise contribution that can be formally written as a combination of the covariances of $\rho$- and $\tau$-statistics: $\mat \Sigma_{\rho}$ and $\mat \Sigma_{\tau}$.

Solving this linear system allows us to estimate the level of systematic which writes:
\begin{align}\label{eq:xi_sys}
    \xi_{\rm PSF, sys}(\vartheta) = 
    \alpha^2 \, &\rho_0(\vartheta)
    + \beta^2 \, \rho_1(\vartheta)
    + \eta^2 \, \rho_3(\vartheta)
    \nonumber\\
    &+ 2\alpha \beta \, \rho_2(\vartheta)
    + 2 \alpha \eta \, \rho_5(\vartheta)
    + 2 \beta \eta \, \rho_4(\vartheta).
\end{align}
This expression represents an additive systematic component to the shear-shear correlation function $\xi^{\gamma \gamma}$. The $\rho$- and $\tau$-statistics thus provide a powerful test of the level of systematics as long as the covariance of the $\tau$-statistics, $\mat \Sigma_\tau$, is accurately estimated. The method to estimate that covariance has to date relied on either simulations or jackknife resampling of the data \citep{zhangGeneralFrameworkRemoving2023,10.1093/mnras/stab918}. The former requires creating many simulations that usually only include galaxy shapes (with the shear following a theoretical power spectrum) and thus does not take into account the contribution from stars and the PSF to the covariance. On the other hand, jackknife resampling is noisy and requires building a sufficiently large number of patches on the sky. In addition, jackknife resampling tends to underestimate the true covariance on large scales \citep{2016MNRAS.456.2662F}. To overcome these limitations and to offer an alternative, we developed in this work a semi-analytical way to compute $\mat \Sigma_\tau$. The analytical expression of $\mat \Sigma_\tau$ will be presented in Sect.~\ref{seq: analytical covariance}.

To solve the linear system \eqref{eq:linear_system_matrix_form}, we use two different methods that both require $\mat \Sigma_\tau$. The first one is a standard Monte-Carlo Markov chain (MCMC) sampling a Gaussian likelihood with covariance $\vec \Sigma_{\tau}$.
The second method solves the following least-square problem:
\begin{equation}
    \vec \Omega^* = \underset{\vec \Omega}{\argmin} \|\vec \tau - \mat R \vec \Omega \|_{\mat \Sigma_\tau}^2,
\end{equation}
This problem has an exact solution that writes:
\begin{equation}
\vec \Omega^* = (\mat R^\textrm{T} \mat \Sigma_{\tau}^{-1} \mat R)^{-1}\mat R^\textrm{T} \vec \tau.   
\end{equation}
To obtain error bars on the parameter estimates from the analytical solution, we sample $\rho$- and $\tau$-statistics from their respective covariances to get a range of solutions of the least-square problem. The covariance of the $\rho$-statistics is computed with jackknife resampling independent of which covariance is used for the $\tau$-statistics. We checked that the results were not sensitive to the variation of the $\rho$-statistics covariance from differences in jackknife patch placement.

In Sect.~\ref{seq:comparison LQ MCMC}
we compare the results obtained using MCMC chains and least-squares minimisation.

\subsection{Semi-analytical covariance}\label{seq: analytical covariance}

\subsubsection{Derivation of the covariance}\label{seq:derivation_covariance}

We develop the covariance for the estimator Eq.~\eqref{eq:hat_xi_pm},
$\hat \xi_+^{ab}$, following \cite{SvWKM02}. We remind the reader that $\hat \xi_+^{ab}$ refers to the two-point correlation function estimator of two spin-2 fields $a$ and $b$. Those fields can represent the ellipticities of galaxies, the PSF, stars, or the difference between the latter two (PSF residuals) in our context.
For the covariance of the three $\tau$-statistics \eqref{eq:tau}, the terms of interest is the covariance $\mat C$ between $\hat\xi_+^{ab}$ and $\hat\xi_+^{dc}$ with $a=d=e$, and $b, c$ being one of the three PSF-related quantities given in Eq.~\eqref{eq:bobs}. We write
\begin{align}
    \MoveEqLeft
  \mat C(\hat \xi_+^{eb}, \hat \xi_+^{ec}; \vartheta_1, \vartheta_2)
  = 
    \left\langle
      \hat \xi_+^{eb}(\vartheta_1) \, \hat \xi_+^{ec}(\vartheta_2)
    \right\rangle
    -
    \left\langle \hat \xi_+^{eb}(\vartheta_1) \right\rangle
    \left\langle \hat \xi_+^{ec}(\vartheta_2) \right\rangle
    \nonumber \\
  = &
    \frac 1 {N^{eb}_\textrm{p}(\vartheta_1) N^{ec}_\textrm{p}(\vartheta_2)}
        \sum_{ijkl} w^e_i w^b_j w^e_k w^c_l
            \left\langle
            \sum_{\alpha=1}^2 e_{i \alpha} b_{j \alpha} 
            \sum_{\beta=1}^2 e_{k \beta} c_{l \beta}
            \right\rangle
    \nonumber \\
    & \quad - \xi_+^{e b}(\vartheta_1) \, \xi_+^{e c}(\vartheta_2)
    \nonumber \\
  = &
  \frac 1 {N^{eb}_\textrm{p}(\vartheta_1) N^{ec}_\textrm{p}(\vartheta_2)}
        \sum_{ijkl} w^e_i w^b_j w^e_k w^c_l
          \sum_{\alpha, \beta=1}^2
          \left\langle
            e_{i \alpha} b_{j \alpha} 
            e_{k \beta} c_{l \beta}
          \right\rangle
    \nonumber \\
    & \quad - \, \xi_+^{e b}(\vartheta_1) \, \xi_+^{e c}(\vartheta_2).
    \label{eq: Covariance}
\end{align}
This expression involves correlations between four ellipticities $\left\langle e_{i \alpha} b_{j \alpha} e_{k \beta} c_{l \beta}\right\rangle$, two of which are galaxy estimates, and the other two are PSF ellipticities or PSF residuals.
We insert Eqs.~\eqref{eq:eobs} and \eqref{eq:bobs} to expand the four-point correlator.
First, odd-power terms in $e^\textrm{s}$, $\gamma$, $b$ and $c$ vanish.
The shot-noise term is zero as well, since $b$ and $c$ are modelled without a stochastic contribution.
The mixed term consists of the galaxy shot noise and the PSF signal contributions, which is of the form
$\left\langle e_{i\alpha}^\textrm{s} e_{k\beta}^\textrm{s} \right\rangle
\left\langle b_{j\alpha} \, c_{l\beta} \right\rangle$. The first factor reduces to $\sigma_e^2 / 2 \times \delta_{\alpha\beta} \delta_{ik}$. The second factor, after carrying out the sum over the ellipticity components, yields $\xi^{bc}_+(\vartheta_{jl})$. In our context, it corresponds to one of the $\rho$-statistics obtained after correlating the fields $b$ and $c$.

The cosmic-variance term can be split into the connected four-point term, and a sum of products of two-point terms according to Wick's theorem.
Here, as in \cite{SvWKM02}, we assume the fields to be Gaussian, and set the connected term to zero.
Tests of this assumption are presented in App.~\ref{seq:gauss_test}, where we use the concept of `transcovariance' matrices \citep{sellentinInsufficiencyArbitrarilyPrecise2018a} to detect non-Gaussianities.
The four-point term is therefore
\begin{align}
    \langle \gamma_{i\alpha}b_{j\alpha}\gamma_{k\beta}c_{l\beta} \rangle
    =
    \langle \gamma_{i \alpha} b_{j\alpha} \rangle \langle \gamma_{k \beta} &c_{l \beta}\rangle + \langle \gamma_{i\alpha}\gamma_{k\beta}\rangle \langle b_{j\alpha}c_{l\beta}\rangle 
    \nonumber \\
    &+ \langle \gamma_{i\alpha} c_{l\beta} \rangle \langle \gamma_{k\beta}b_{j\alpha}\rangle,
    \label{eq:wick}
\end{align}
with $\alpha, \beta \in \{1, 2\}$, $i \ne j$ and $k \ne l$.
The first term on the right-hand side of Eq.~\eqref{eq:wick} separates into sums over $(i, j, \alpha)$ and $(k, l, \beta)$, the result of which is $\xi_+^{\gamma b}(\vartheta_1) \xi_+^{\gamma c}(\vartheta_2)$. This product of the means is subtracted in Eq.~\eqref{eq: Covariance}.
The remaining terms lead, using Eq.~\eqref{eq:terms_ab}, to:
\begin{align}
    \MoveEqLeft
    \sum_{\alpha, \beta=1}^2
        \left[
        \left\langle \gamma_{i \alpha} \gamma_{k \beta} \right\rangle
        \left\langle b_{j \alpha} c_{l \beta} \right\rangle
        + \left\langle \gamma_{i \alpha} c_{l \beta} \right\rangle
        \left\langle \gamma_{k \beta} b_{j \alpha} \right\rangle
        \right]
    \nonumber \\
    = & \,
     \frac{1}{2} \left[
     \xi_+^{\gamma \gamma}(ik)\xi_+^{bc}(jl)+\xi_+^{\gamma c}(il)\xi_+^{\gamma b}(jk) \right.
     \nonumber \\
     & \quad +\xi_-^{\gamma \gamma}(ik)\xi_-^{bc}(jl)\cos(4(\phi_{ik}-\phi_{jl}))
    \nonumber \\
    & \quad +\left. \xi_-^{\gamma c}(il) \xi_-^{\gamma b}(jk)\cos(4(\phi_{il}-\phi_{jk}))
    \right].
\end{align}

The total covariance is then
\begin{align}
\MoveEqLeft
  \mat C(\hat \xi_+^{eb}, \hat \xi_+^{ec}; \vartheta_1, \vartheta_2)
  =
    \frac 1 {N_\textrm{p}^{\rm eb}(\vartheta_1) N_\textrm{p}^{\rm ec}(\vartheta_2)}
    \nonumber \\
    &\times \Bigg\{\frac{\sigma_e^2} 2 \sum_{ijk} (w^e_i)^2 w^b_j w^c_k
      \Delta_{\vartheta_1}(\vartheta_{ij})
      \Delta_{\vartheta_2}(\vartheta_{ik})
        \xi^{bc}_+(\vartheta_{jk})
    \nonumber \\
      &
      +
      \frac 1 2 \sum_{ijkl} w^e_i w^b_j w^e_k w^c_l
      \Delta_{\vartheta_1}(\vartheta_{ij})
      \Delta_{\vartheta_2}(\vartheta_{kl})
      \nonumber \\
      & \quad \times \left[
      \left(
        \xi_+^{\gamma\gamma}(\vartheta_{il}) \xi_+^{bc}(\vartheta_{jk})
        +  \xi_+^{\gamma c}(\vartheta_{il}) \xi_+^{\gamma b}(\vartheta_{jk})
        \right)
      \right.
      \nonumber\\
      &\quad +
      \left.
      \left(
      \xi_-^{\gamma \gamma}(\vartheta_{il}) \xi_-^{bc}(\vartheta_{jk})
      +
      \xi_-^{\gamma c}(\vartheta_{il}) \xi_-^{\gamma b}(\vartheta_{jk})
      \right)
      \cos 4 \left( \phi_{il} - \phi_{jk} \right)
      \right]
    \Bigg\}
      .
  \label{eq:cov_abcd}
\end{align}
For the cosmic-variance term we made use of the invariance of each summand under exchange of the indices $k$ and $l$. As discussed earlier, since we did not add a stochastic element $b^{\rm s}$ to the PSF fields the mixed-term is $1/4$ of that of  \cite{SvWKM02}, and the pure shot noise term vanishes. However, such a shot noise that is measured in actual data cannot be neglected and a caveat of our model is the absence of shot noise. We will describe in Sect.~\ref{seq:numerical_computation} how we take into account the measured shot noise in practice.

\subsubsection{Ensemble averages}

Given a catalog containing the different fields involved in the computation of the covariance \eqref{eq:cov_abcd}, the latter can be calculated using galaxy positions and their weights. However, the sum over three or even four galaxy positions is not easily tractable given the amount of galaxies in our catalogs. A Monte-Carlo approach that drew random simulated galaxy positions distributed uniformly over the survey footprint was developed in \cite{KS04}. Here, we follow \cite{SvWKM02} and replace the sums over galaxy positions by ensemble averages. As in \cite{SvWKM02} we set all weights to unity and consider a survey geometry of solid angle $A$ and galaxy number density $n^a$ for a field $a$. The number of pairs is approximated as
\begin{align}
    N_\textrm{p}^{ab}(\vartheta) \approx 2 \pi \Delta\vartheta \vartheta n^a n^b A.
\end{align}

The ensemble average for one galaxy corresponds to the area integral over the survey footprint normalised by the survey area. For $N$ galaxies the ensemble averages for all galaxies are multiplied, resulting in the operator
\begin{equation}
    E = \prod_{i=1}^N \left( \frac{1}{A} \int_A \textrm{d}^2 \vec \theta_i \right).
    \label{eq:E}
\end{equation}
For the mixed term $M_e$, we take the ensemble average over triplets of object positions.
The three catalogues have numbers of objects $N^e$, $N^b$, and $N^c$, respectively.
All but three integrals in \eqref{eq:E} are trivial and reduce to unity. The sum of triple integrals can be written as $N^e N^b N^c$ permutations of the same expression. With all prefactors, this is then
\begin{align}
    E\left[ M_e \right] = &
    \frac{1}{N_\textrm{p}^{eb}(\vartheta_1) N_\textrm{p}^{ec}(\vartheta_2)}
    \frac{\sigma_e^2} 2 \frac{N^e N^b N^c}{A^3}
    \nonumber \\
    & \times \int \textrm{d}^2 \theta_1 \int \textrm{d}^2 \theta_2 \int \textrm{d}^2 \theta_3 \,
    \Delta_{\vartheta_1} (\theta_{12})
    \Delta_{\vartheta_2} (\theta_{13})
    \, \xi_+^{bc}\left( \theta_{23} \right),
\end{align}
where $\theta_{ij} = |\vec \theta_i - \vec \theta_j |$.
We substitute $\vec \theta_2^\prime = \vec \theta_2 - \vec \theta_1$
and $\vec \theta_3^\prime = \vec \theta_3 - \vec \theta_1$;
the $\vec \theta_1$-integration can then be carried out trivially to yield the area $A$.

Conveniently, we split the area integrals into radial and azimuthal parts.
The bin indicator functions restrict $\theta_2^\prime$ (resp.~$\theta_3^\prime$) in bins of width $\Delta \theta$ around $\vartheta_1$ (resp.~$\vartheta_2$). Assuming that
$\xi_+^{bc}$ does not vary much over the bin width, we can solve the two radial integrals.
This leads to the intermediate result
\begin{align}
   E\left[ M_e \right] = &
    \frac{1}{N_\textrm{p}^{eb}(\vartheta_1) N_\textrm{p}^{ec}(\vartheta_2)}
    \frac{\sigma_e^2} 2 \frac{N^e N^b N^c}{A^2}
    \Delta \vartheta^2 \vartheta_1 \vartheta_2
    \nonumber \\
    & \times
    \int\limits_0^{2\pi} \textrm{d} \varphi_1
    \int\limits_0^{2\pi} \textrm{d} \varphi_2 \,
    \, \xi_+^{bc}\left( \sqrt{ \vartheta_1 + \vartheta_2 - 2 \vartheta_1 \vartheta_2 \cos (\varphi_2 - \varphi_1) } \right) .
\end{align}
One of the angular integrals can be carried out trivially to yield $2 \pi$. Expanding the number of pairs, we find, 
analogously to \cite{SvWKM02},
\begin{align}\label{eq: mixed term}
    E\left[ M_e \right] =
        \frac{\sigma_e^2}{2 \pi A n^e}
        \int\limits_0^\pi \textrm{d} \varphi \,
        \xi_+^{bc}\left( \sqrt{ \vartheta_1^2 + \vartheta_2^2 - 2 \vartheta_1 \vartheta_2 \cos \varphi}
        \right).
\end{align}
Note the reduced integration range $[0; \pi]$ due to the periodicity of the integrand. As expected, this is one fourth of the corresponding term Eq.~(32) in \cite{SvWKM02}.

The cosmic-variance term is  computed in a similar way as the mixed term. Under the ensemble average operator Eq.~\eqref{eq:E}, these  can be written as $\approx N^e N^e N^b N^c$ permutations of a product over four galaxy positions. All three addends of the cosmic-variance term
can be written in the form
\begin{align}
\MoveEqLeft
   E\left[ V^\textrm{term} \right] =
    \frac{1}{N_\textrm{p}^{eb}(\vartheta_1) N_\textrm{p}^{ec}(\vartheta_2)}
    \frac{\left(N^e\right)^2 N^b N^c}{2 A^4}
    \int \textrm{d}^2 \theta_1
    \int \textrm{d}^2 \theta_2
    \, \Delta_{\vartheta_1} (\theta_{12})
    \nonumber \\
    & \times
    \int \textrm{d}^2 \theta_3
    \int \textrm{d}^2 \theta_4
    \, \Delta_{\vartheta_2} (\theta_{34})
    \, F_1(\vec \theta_{23})
    \, F_2(\vec \theta_{14}).
\end{align}
The first two terms depend on the scalars $\theta_{23}$ and $\theta_{14}$. For the third term, we can expand the cosine function into products of cosines and sines, and thus the above separation into $F_1$ and $F_2$ is valid.

We perform the variable substitutions $\vec \phi_1 = \vec \theta_2 - \vec \theta_1$ and $\vec \phi_2 = \vec \theta_4 - \vec \theta_3$. Since $\vec \theta_{23} = \vec \theta_3 - \vec \theta_1 - \vec \phi_1$ and $\vec \theta_4 - \vec \theta_1 = \vec \theta_3 - \vec \theta_1 + \phi_2$, the arguments of $F_1$ and $F_2$ only depend on $\vec \phi \equiv \vec \theta_1 - \vec \theta_3$. Therefore, the $\vec \theta_3$-integration can be carried out trivially to yield a factor $A$. We split up the integrals over the bin indicator arguments into radial and azimuthal part, and find
\begin{align}
\MoveEqLeft
   E\left[ V^\textrm{term} \right] =
    \frac{1}{N_\textrm{p}^{eb}(\vartheta_1) N_\textrm{p}^{ec}(\vartheta_2)}
    \frac{\left(N^e\right)^2 N^b N^c}{2 A^3}
    \int \textrm{d}^2 \phi
    \int \textrm{d} \phi_1 \, \phi_1
    \Delta_{\vartheta_1}(\phi_1)
    \nonumber \\ & \times
    \int \textrm{d} \varphi_1
    \int \textrm{d} \phi_2 \, \phi_2
    \Delta_{\vartheta_2}(\phi_2)
    \int \textrm{d} \varphi_2
    \, F_1(\vec \phi - \vec \phi_1)
    \, F_2(\vec \phi + \vec \phi_2).
\end{align}
Here, we have defined $\varphi_i$ as the polar angles of $\vec \phi_i$ for $i=1, 2$.
Using the bin indicator functions the radial integrals over $\phi_1$ and $\phi_2$ can be simplified, under the assumption that the integrand varies slowly over the bin width, see above. At the same time we replace the vectors $\vec \phi_i$ by $(\vartheta_i, \varphi_i), i=1, 2$.
%
The ensemble average for any of the terms is
\begin{align}\label{eq:cosmic_var_cov}
  E\left[ V^\textrm{term} \right]
   = &
    \frac 1 {8 \pi^2 A}
    \int\limits_0^\infty \textrm{d} \phi \, \phi 
    \int\limits_0^{2\pi}  \textrm{d} \varphi
    \nonumber \\ & \times 
    \int\limits_0^{2\pi}  \textrm{d} \varphi_1
        F_1(\vec \phi - \vec \vartheta_1) \,
    \int\limits_0^{2\pi}  \textrm{d} \varphi_2
        F_2(\vec \phi + \vec \vartheta_2) .
\end{align}
%
We rewrite the arguments of the functions $F_i$ as
    \begin{equation}
        \vec \psi_1 = \vec \phi - \vec \vartheta_1 = \psi_1 e^{i\varphi_{\psi_1}};
        \quad
        \vec \psi_2 = \vec \phi + \vec \vartheta_2 = \psi_2 e^{i\varphi_{\psi_2}},
    \end{equation}
where we identify a vector with its complex notation to make apparent the modulus and argument of $\vec \psi_1$ and $\vec \psi_2$.

We need to consider two cases for the functions $F_1$ and $F_2$.
%
    The first case corresponds to the product of the `$+$'-correlators, the first two terms of the cosmic-variance covariance.
    The arguments of the functions $F_i$ are scalar variables,
    $F_1(\vec \psi_1) = \xi_+^{\rm ab}(\psi_1)$
    and
    $F_2(\vec \psi_2) = \xi_+^{\rm ab}(\psi_2)$
    which depend only on the difference between the polar angle $\varphi - \varphi_1$ and $\varphi-\varphi_2$. It allows to change variables and carry the outer integral over $\varphi$. The resulting term can then be written, using periodicity and parity arguments, as:
    \begin{equation}\label{eq:cosmic_variance+}
        E[V^{\rm term}_+] = \frac{1}{\pi A}\int\limits_0^\infty \textrm{d} \phi \, \phi 
    \int\limits_0^{\pi}  \textrm{d} \varphi_1
        \xi_+^{\rm ab}(\psi_1) \,
    \int\limits_0^{\pi}  \textrm{d} \varphi_2
        \xi_+^{\rm cd}(\psi_2) .
    \end{equation}
    %
    where $a$, $b$, $c$ and $d$ will be replaced accordingly to get the `+'-cosmic-variance term in Eq.~\eqref{eq: Covariance}.
    
    The `-'-component of the cosmic variance term in Eq.~\eqref{eq: Covariance} is obtained by using in Eq.~\eqref{eq:cosmic_var_cov} the following definitions:
    \begin{align}
        F_1(\vec \psi_1) &= \xi_-^{\rm ab}(\psi_1) \, \textrm{cs} 4\varphi_{\psi_1};
        \quad
        F_2(\vec \psi_2) = \xi_-^{\rm cd}(\psi_2) \, \textrm{cs} 4\varphi_{\psi_2},
    \end{align}
    with $\textrm{cs} \in \{\cos, \sin \}$. Contrary to the first (`+') case, the 4-dimensional integral cannot be simplified with a change of variable as the polar angles of $\vec \psi_1$ and $\vec \psi_2$ do not depend only on the difference $\varphi - \varphi_i$ with $i=1,2$. Note that this is incorrectly stated in \cite{SvWKM02}.
    One thus has to carry out the full 4-dimensional integral using the following expression:
%
%
\begin{align}\label{eq: cosmic-variance term}
  E\left[ V^\textrm{term}_- \right]
  =
    &\frac 1 {8 \pi^2 A}  
    \int\limits_0^\infty \textrm{d} \phi \, \phi  \,
    \int\limits_0^{2\pi} \textrm{d} \varphi \, \nonumber\\
    &\times \int\limits_0^{2\pi}  \textrm{d} \varphi_1
        \xi_-^{\rm ab}(\psi_1) \rm{cs} 4\varphi_{\psi_1} \,
    \int\limits_0^{2\pi}  \textrm{d} \varphi_2
        \xi_-^{\rm cd}(\psi_2) \rm{cs} 4\varphi_{\psi_2}.
\end{align}

With all those terms, the covariance of the $\tau_i$ is constructed using Eq.~\eqref{eq:cov_abcd}, the mixed term $M_e$ Eq.~\eqref{eq: mixed term} and the cosmic-variance terms Eqs.~\eqref{eq:cosmic_variance+} and \eqref{eq: cosmic-variance term}. For example for $\tau_0$ we set      
$a = c = \gamma, b = d = \ep$. The covariance matrix depends on the $\rho$-statistics, the $\tau$-statistics and the shear-shear correlation functions integrated on various angular ranges. The following section describe how the computation of the semi-analytical covariance matrix is carried out. In what follows, we will refer to the `+'-contribution to the cosmic variance coming from $\xi_+^{\gamma\gamma}(\vartheta_{il}) \xi_+^{bc}(\vartheta_{jk})$ in \eqref{eq:cov_abcd} as the $\rho_+$ contribution and $\xi_+^{\gamma c}(\vartheta_{il}) \xi_+^{\gamma b}(\vartheta_{jk})$ as the $\tau_+$ (See Fig.\ref{fig:diag_cov}).

\subsubsection{Numerical computation}\label{seq:numerical_computation}

In practice, to compute the integrals in Eqs.~\eqref{eq: mixed term}, \eqref{eq:cosmic_variance+}, and \eqref{eq: cosmic-variance term}, we 
first
use \textsc{TreeCorr}\footnote{Link to \href{ https://rmjarvis.github.io/TreeCorr/\_build/html/index.html}{TreeCorr documentation.}} \citep{JBJ04} to estimate the correlation functions. These are calculated from the observed galaxy, star, and PSF catalogues on a large angular range, and interpolate between those points using interpolators from \textsc{SciPy}\footnote{Link to \href{https://scipy.org/}{\textsc{SciPy} documentation}} \citep{virtanenSciPyFundamentalAlgorithms2020}. The covariance is thus semi-analytical as we have no theoretical predictions for the $\rho$- and $\tau$-statistics and use the statistics estimated on the data to build the covariance matrix. Due to the assumptions made previously, there is no explicit shot noise contribution in our semi-analytical covariance model. Using an expression similar to Eq.~(27) of \cite{SvWKM02} does not recover the correct shot noise consistent with the observations. To accurately compute the shot noise contribution of the data to the covariance matrix, we use \textsc{TreeCorr} that allows us to approximate the shot noise contribution when computing two-point correlation functions. We verified that the shot noise is accurately recovered by comparing it with the diagonal of the covariance matrix on small scales obtained with simulations and jackknife resampling (see Fig.~\ref{fig:diag_cov}). We underline that, in this work, we restricted ourselves to the covariance of the `$+$' component of the $\tau$-statistics, as the `$-$' component is noisy and does not provide a significant improvement in the constraints on the parameters $\vec \Omega$. However, the formalism introduced in Section \ref{seq: analytical covariance} allows the interested reader to derive an analytical expression for the auto-correlation of the `$-$' component, or the cross-correlation between the `$+$' and `$-$' components for the $\tau$-statistics.

In what follows, we compute the $\rho$- and $\tau$-statistics on $20$ angular bins between $0.1$ and $250$ arcmin. \textsc{TreeCorr}, as a treecode, provides a fast computation of correlation functions. We use MCMC and least-squares to estimate the parameters in Eq.~\eqref{eq:linear_system}. The Monte-Carlo chains are run using \textsc{emcee} \citep{foreman-mackeyEmceeMCMCHammer2013} with $124$ walkers and performing $10,000$ steps each. As mentioned earlier, with this approach, the covariance of the $\rho$-statistics is not taken into account: The $\rho$-statistics used to obtain the $\tau$-statistics given parameters $\vec \Omega$ are fixed to the estimated values computed via \textsc{TreeCorr}.
To compute the uncertainty of the least-square solution however, we sample both the $\rho$- and the $\tau$-statistics covariance. We also assume that the likelihood is Gaussian and we provide tests of this assumption in App.~\ref{seq:gauss_test} using the so-called `transcovariance' matrices. Indeed, at intermediate scales, the Central Limit Theorem (CLT) ensures that the data is Gaussian distributed but there is no guarantee on large and small scales. On small scales, non-linear transformations of the field can introduce non-Gaussianities in the data. On large scales, the two-point correlation function of two Gaussian fields is not Gaussian distributed and poorly sampled and could thus introduce non-Gaussianities. The test discussed in App.~\ref{seq:gauss_test} does not detect significant non-Gaussianities in the data. Hence, a Gaussian likelihood is a reasonable assumption.

\subsubsection{Jackknife resampling}\label{seq:jacknife_resampling}

\textsc{TreeCorr} also allows us to compute covariances using jackknife resampling. This consists in dividing the sky into $N_{\rm patch}$ patches and computing $N_{\rm patch}$ correlation functions, where a patch is removed from the footprint for each computation. \textsc{TreeCorr} partitions the sky into patches of roughly equal area using a $K$-means clustering algorithm.
We choose the rather low number of patches $N_{\rm patch}=150$ to avoid biases of the covariance estimate due to patches containing no objects:
Patches are created based on either the star or galaxy sample, but used jointly for both samples in the cross-correlations. Empty patches occur due to the differences in footprint and density of stars and galaxies (e.g., in regions of high star density no  galaxies are detected), which is less likely to happen for large patch sizes. The choice of the number of patches is also aimed at mitigating the fluctuation of the number of stars in each patch that is more important in small patches.
However, because of the relatively small number of patches, the covariance is sensitive to the patching, which has some stochasticity due to the random initialisation. Therefore, we repeat each jackknife resampling $100$ times with different initialisations and take the mean covariance to marginalize over the patching of the sky. We find that this procedure, albeit time-consuming, provides significantly more stable results.

\section{Data and simulations}\label{seq: data and simulations}

\begin{figure*}
    \centering
    \includegraphics[width=0.9\linewidth]{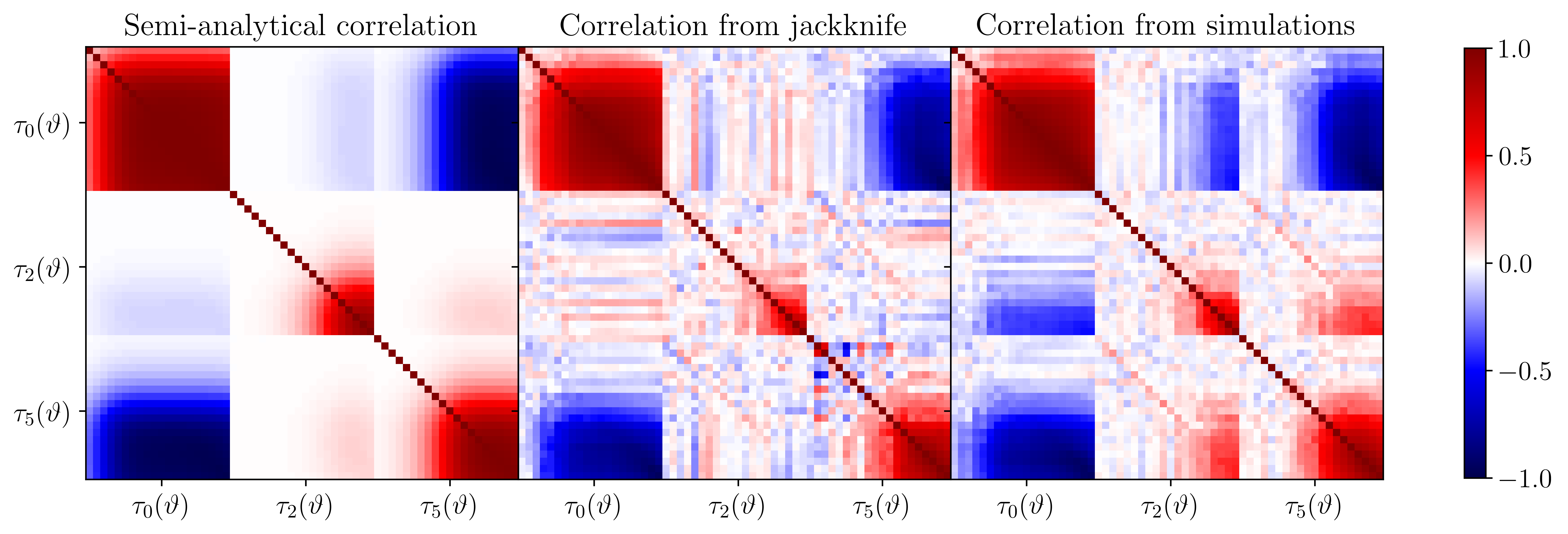}
    \caption{Correlation matrices of the $\tau$-statistics for different methods. The panels from left to right correspond to the analytical expressions from Eq.~\eqref{eq: Covariance}, jackknife resampling, and simulations, respectively.}
    \label{fig:corr_matrix_tau}
\end{figure*}

\begin{figure*}
    \centering
    \includegraphics[width=\linewidth]{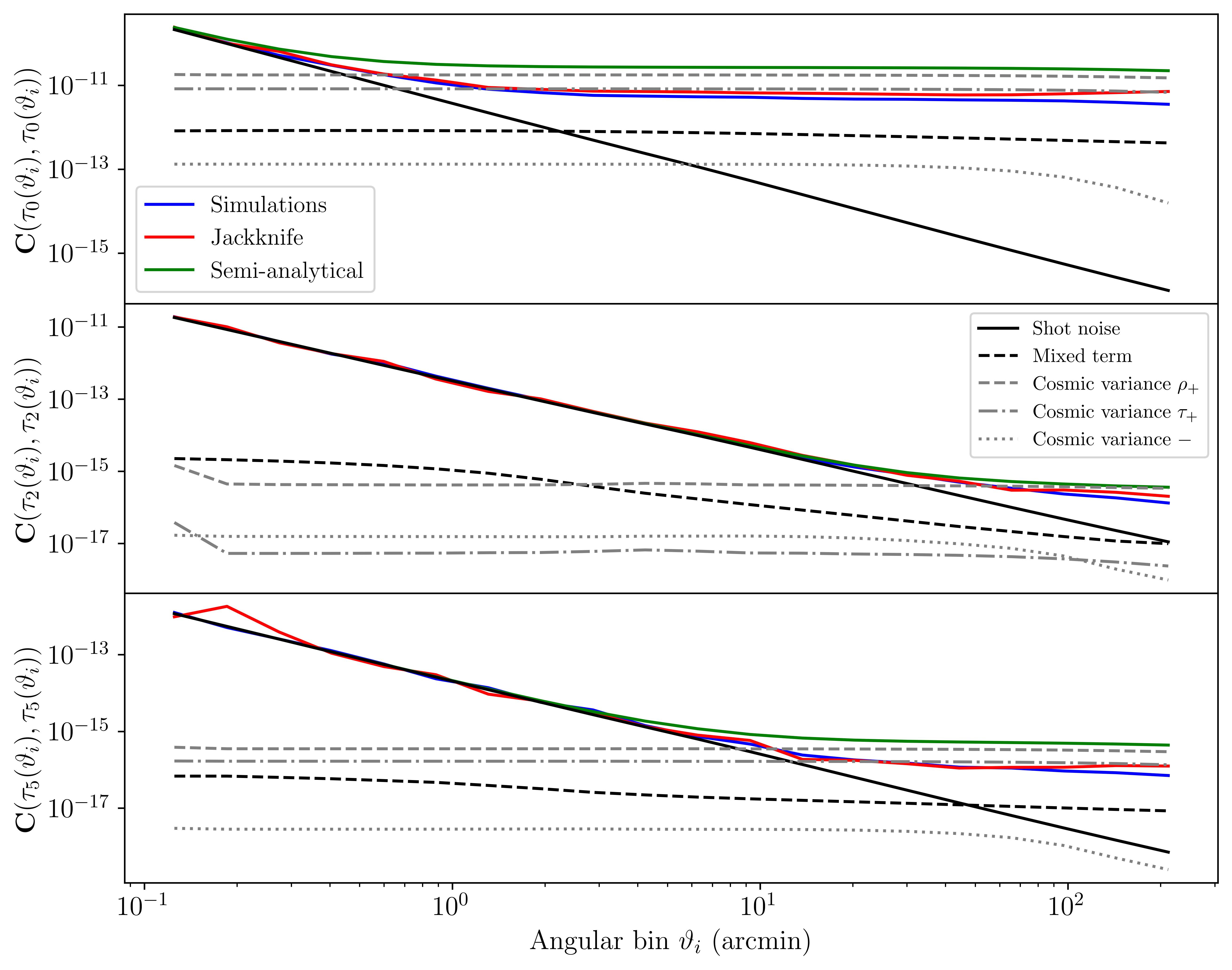}
    \caption{Diagonal component of the three $\tau$-statistics covariances for the different methods used to estimate the covariance. \textit{Upper panel}: Diagonal of the $\tau_0$-$\tau_0$ covariance. \textit{Middle panel}: Diagonal of the $\tau_2$-$\tau_2$ covariance. \textit{Lower panel}: Diagonal of the $\tau_5$-$\tau_5$ covariance. Black and grey lines correspond to individual contribution of each term in the covariance in Eq.~\eqref{eq: Covariance}. Small scales are shot noise-dominated. We see a good agreement between the three methods, and in particular, we validate the shot noise estimate of the semi-analytical covariance obtained from \textsc{TreeCorr} (See Section \ref{seq:derivation_covariance}).} 
    \label{fig:diag_cov}
\end{figure*} 

\begin{figure*}
    \centering
    \includegraphics[width=0.45\linewidth]{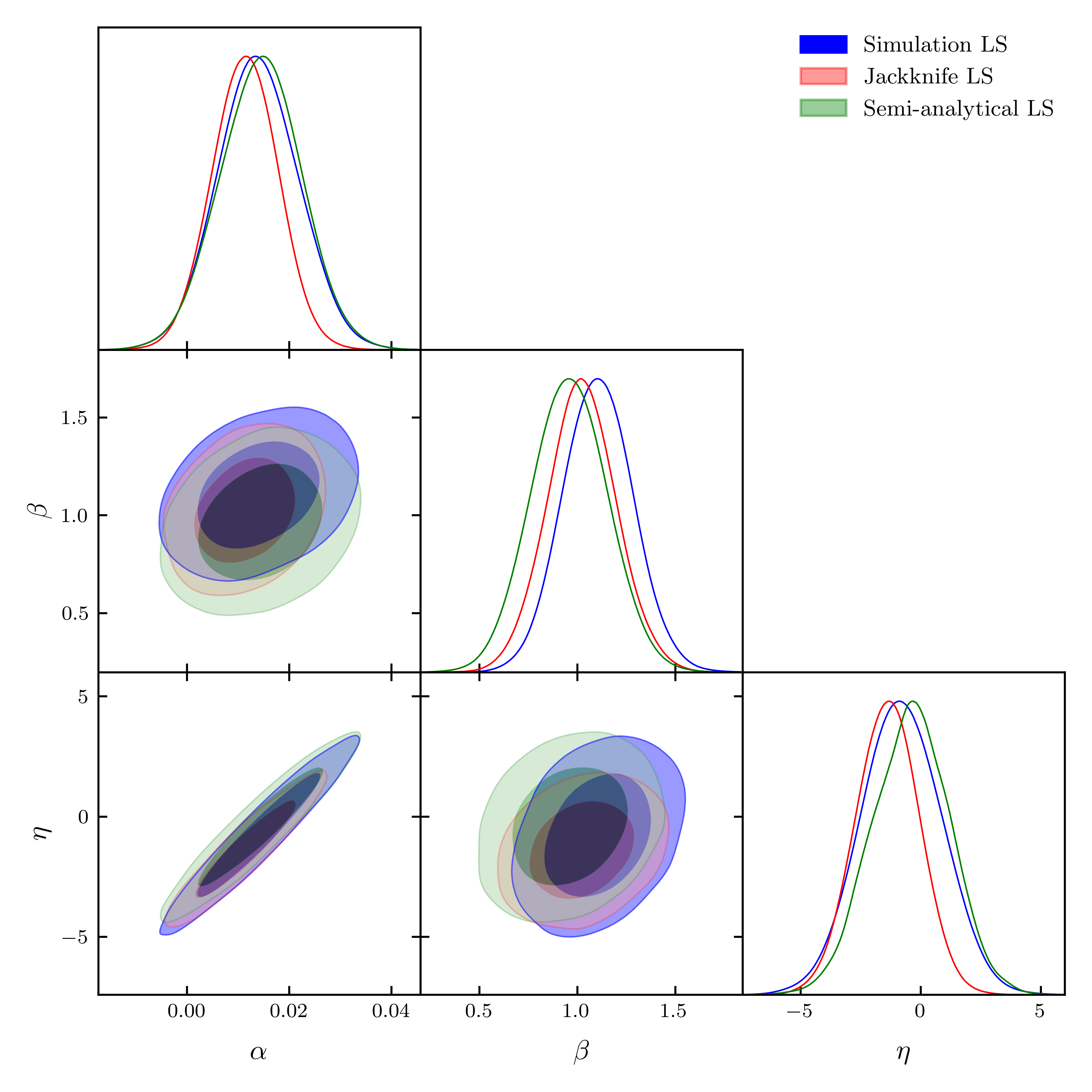}
    \includegraphics[width=0.45\linewidth]{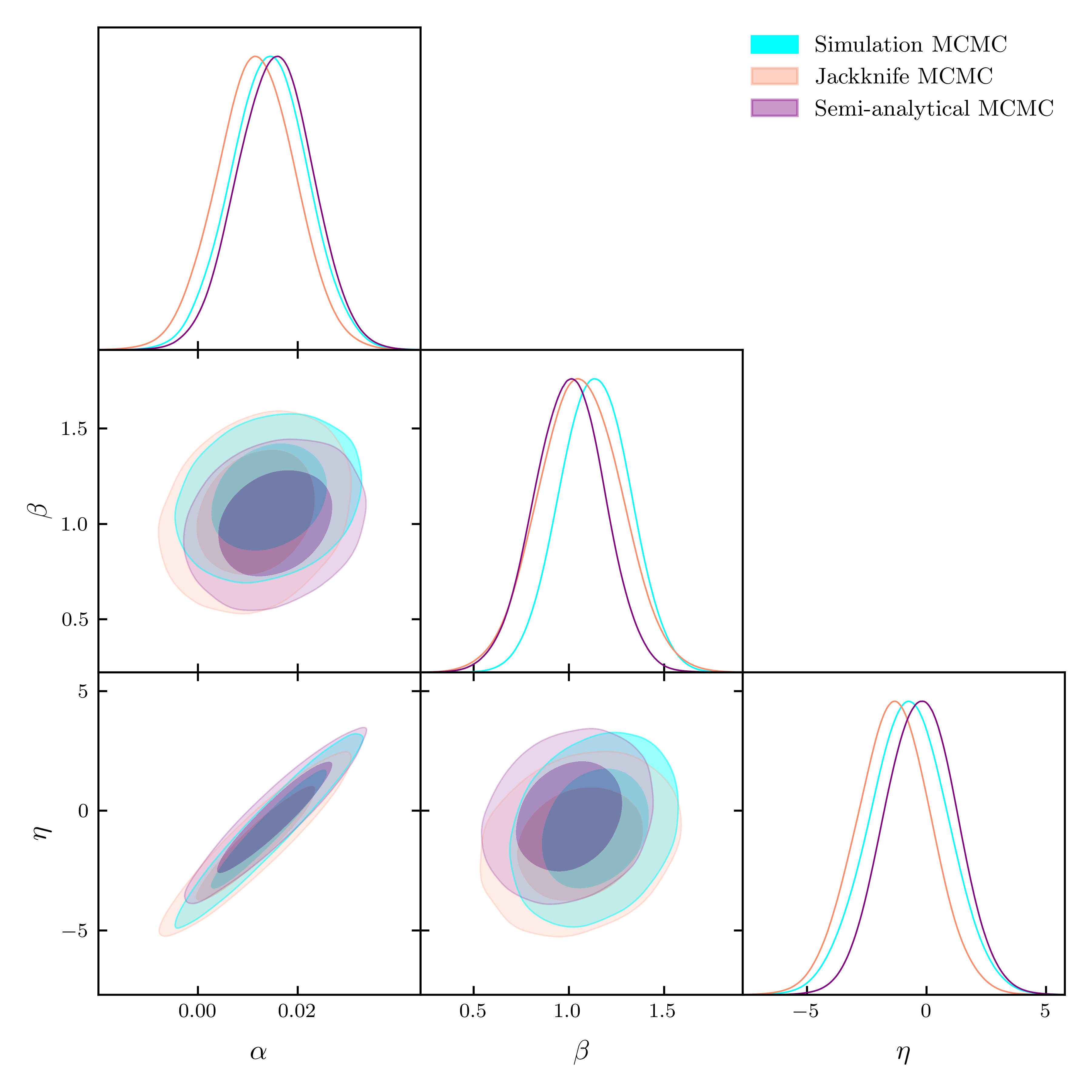}
    \caption{Constraints obtained on parameters $\vec \Omega = (\alpha, \beta, \eta)^\textrm{T}$ of the PSF error model. \textit{Left panel:} Constraints are obtained using the least-square method. \textit{Right panel:} Constraints are obtained using MCMC.}
    \label{fig:contours_lq_sp_v1.4}
\end{figure*}

\begin{table*}\label{tab:parameter values}
    \caption{Summary of the constraints obtained on $\alpha$, $\beta$ and $\eta$ for the different sampling methods and covariance estimates. The errors are given at the 68\% confidence level. The reduced chi-square is given for a number of degrees of freedom of 57 (60 angular bins in total and 3 parameters).}
    \centering
        \begin{tabular}{ccccccc}
            Covariance estimate & Sampling method & $\alpha$ & $\beta$ & $\eta$ & $\chi^2$ & $\chi^2_{\rm red}$\\ \hline
            Semi-Analytical & Least-Squares & $0.014^{+0.008}_{-0.008}$ & $0.96^{+0.19}_{-0.19}$ & $-0.389^{+1.6}_{-1.7}$ & $93$ & $1.63$\\
            Semi-Analytical & MCMC & $0.016^{+0.007}_{-0.008}$ & $1.0^{+0.18}_{-0.19}$ & $-0.243^{+1.5}_{-1.5}$ & $93$ & $1.63$\\
            Jackknife & Least-Squares & $0.011^{+0.006}_{-0.006}$ & $1.0^{+0.18}_{-0.18}$ & $-1.4^{+1.3}_{-1.3}$ & $49$ & $0.86$\\
            Jackknife & MCMC & $0.012^{+0.008}_{-0.008}$ & $1.1^{+0.22}_{-0.21}$ & $-1.4^{+1.5}_{-1.5}$ & $49$ & $0.86$\\
            Simulation & Least-Squares& $0.014^{+0.008}_{-0.008}$ & $1.1^{+0.18}_{-0.18}$ & $-0.81^{+1.7}_{-1.7}$ & $67$ & $1.18$\\
            Simulation & MCMC & $0.014^{+0.008}_{-0.008}$ & $1.1^{+0.18}_{-0.18}$ & $-0.75^{+1.6}_{-1.7}$ & $68$ & $1.19$\\
        \end{tabular}
\end{table*}

\begin{figure*}
    \centering
    \includegraphics[width=0.8\linewidth]{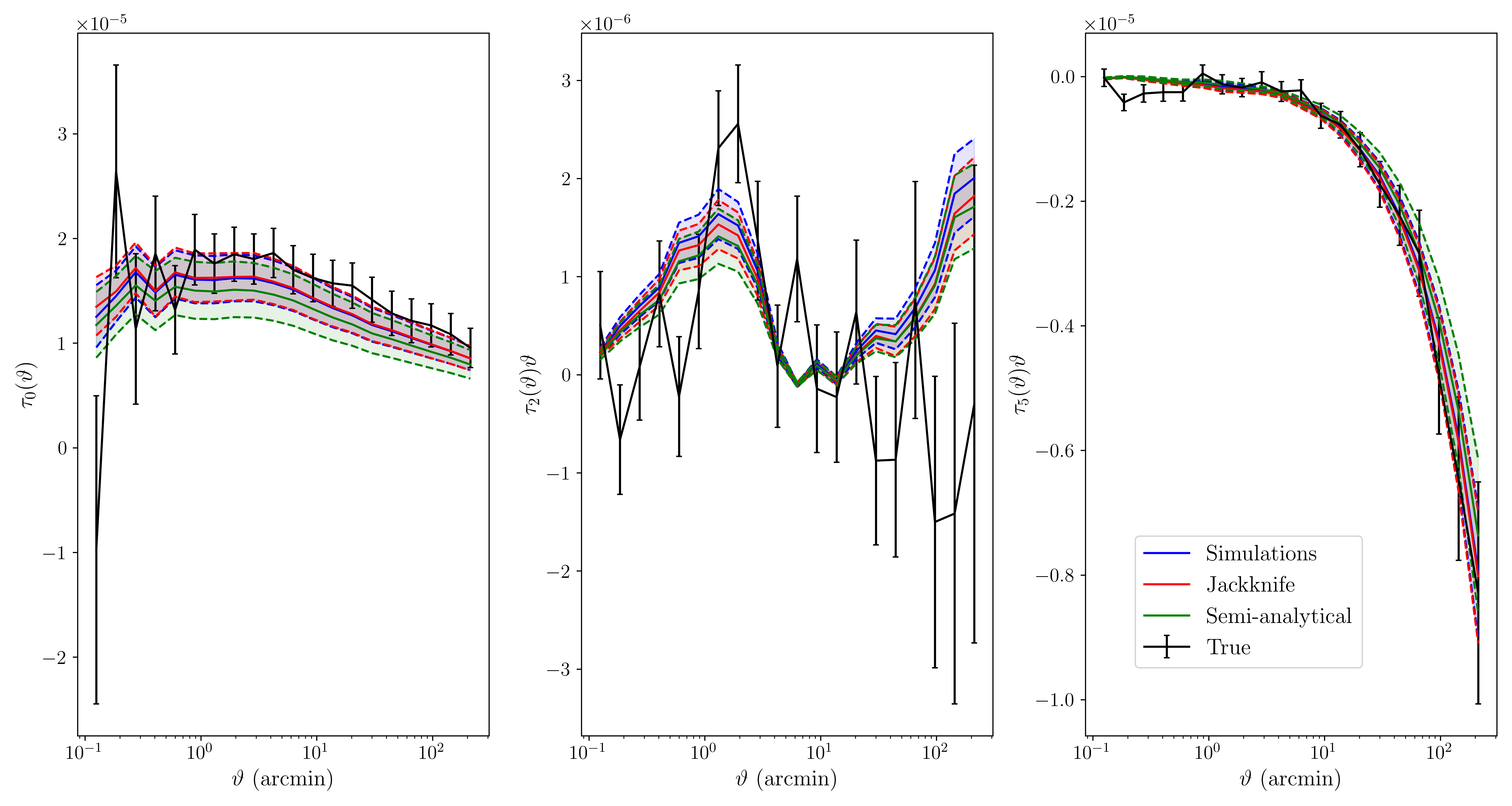}
    \caption{Observed data with error bars and best-fit of the $\tau$-statistics at the 68\% confidence level for the three different methods used to estimate the covariance matrix of the $\tau$-statistics. The $\tau$-statistics are estimated using parameters $(\alpha, \beta, \eta)$ sampled with least-squares. Note that $\tau_2$ and $\tau_5$ are multiplied by $\vartheta$ in the middle and right panels.}
    \label{fig:best_fit}
\end{figure*}

\begin{figure*}
    \centering
    \includegraphics[width=0.8\linewidth]{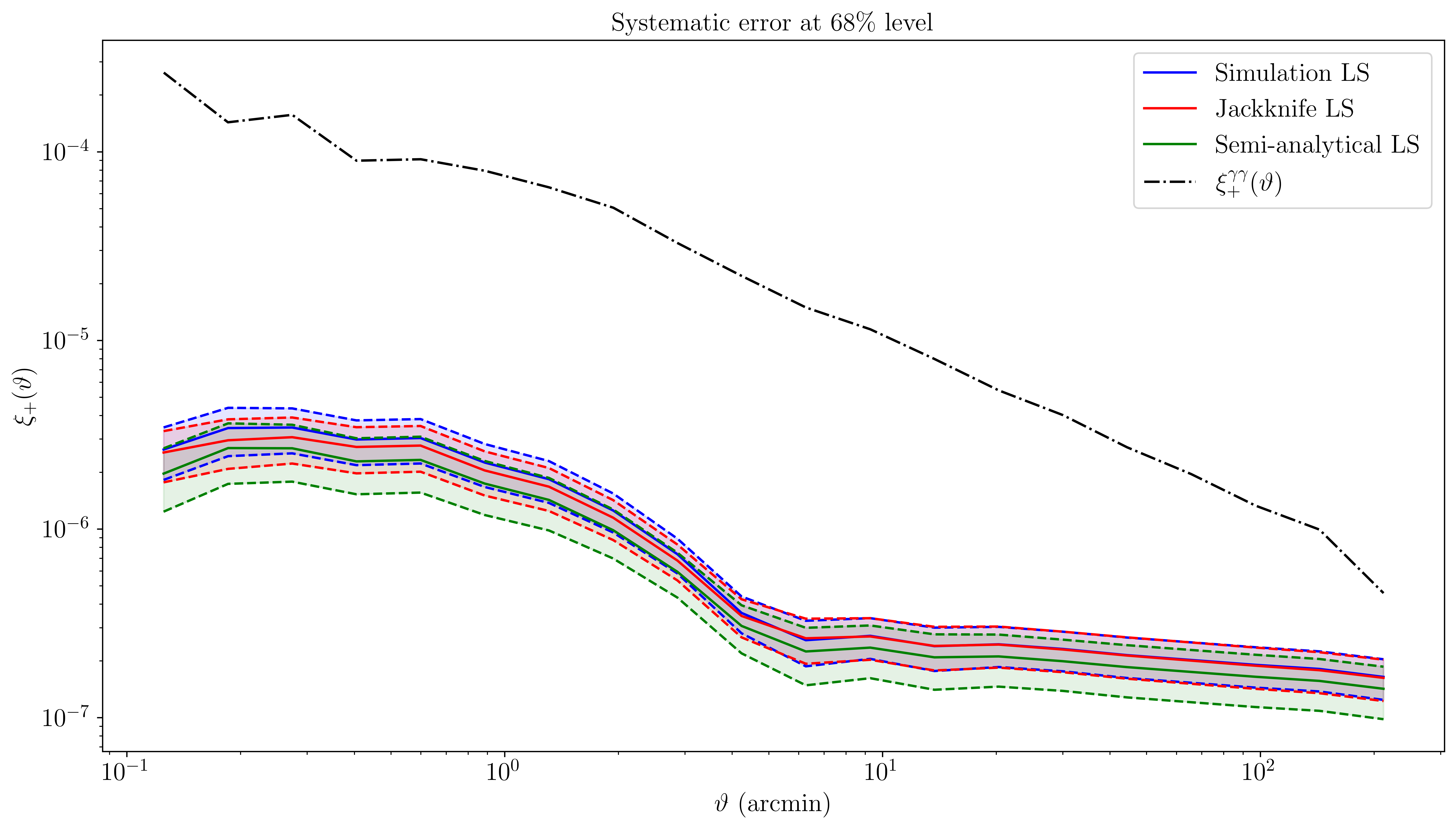}
    \includegraphics[width=0.8\linewidth]{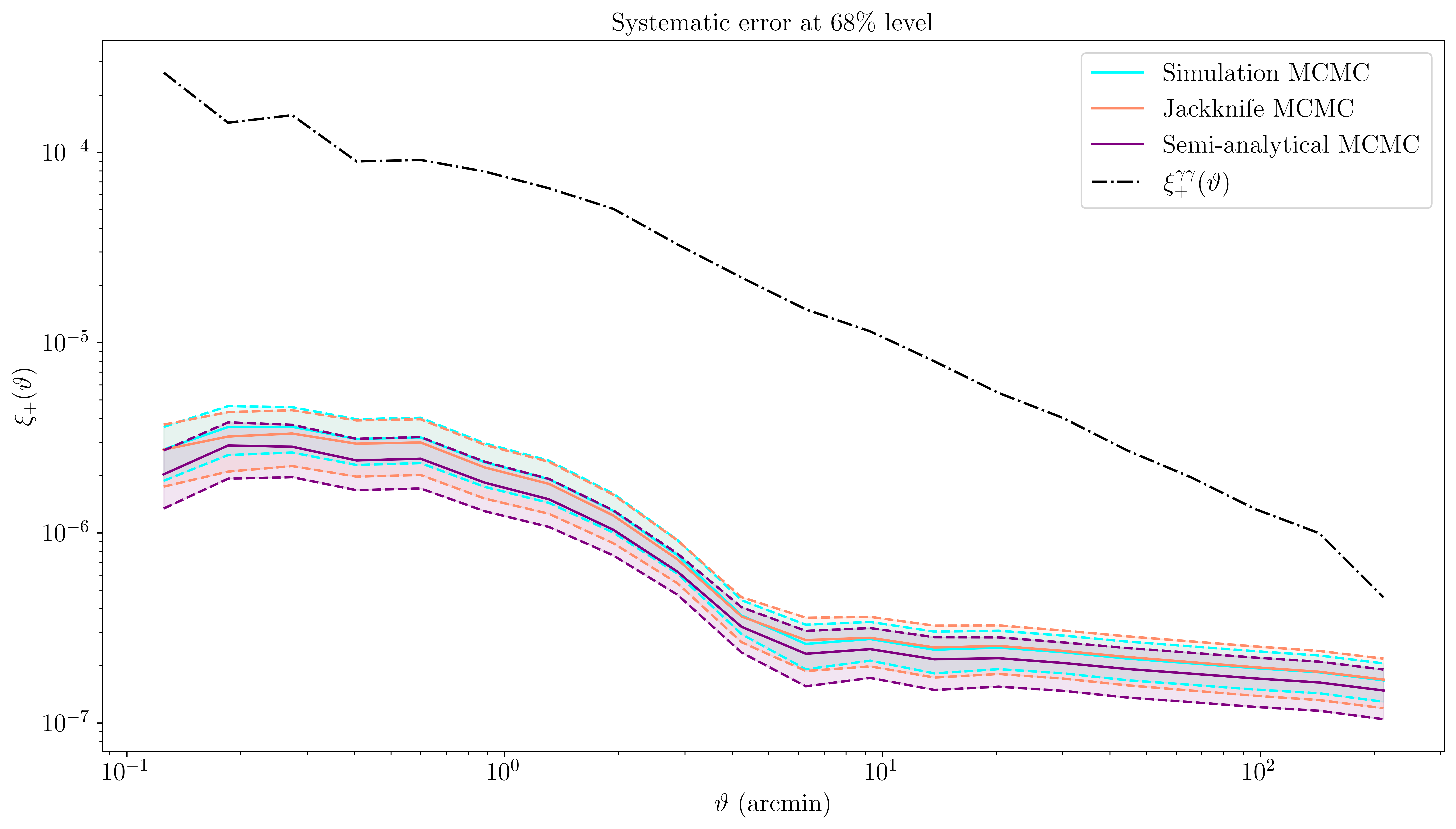}
    \caption{Level of systematic error for the different method to estimate the covariance matrices. We show the confidence interval for the level of systematic at the 68\% confidence level. \textit{Top Panel:} Using least-squares to estimate the parameters. \textit{Bottom panel:} Using MCMC to estimate the parameters. The dotted lines correspond to the $\xi_{+}$ two-point correlation function with respect to which the level of systematics has to be compared.}
    \label{fig:systematic error level}
\end{figure*}

\begin{figure*}
    \centering
    \includegraphics[width=\linewidth]{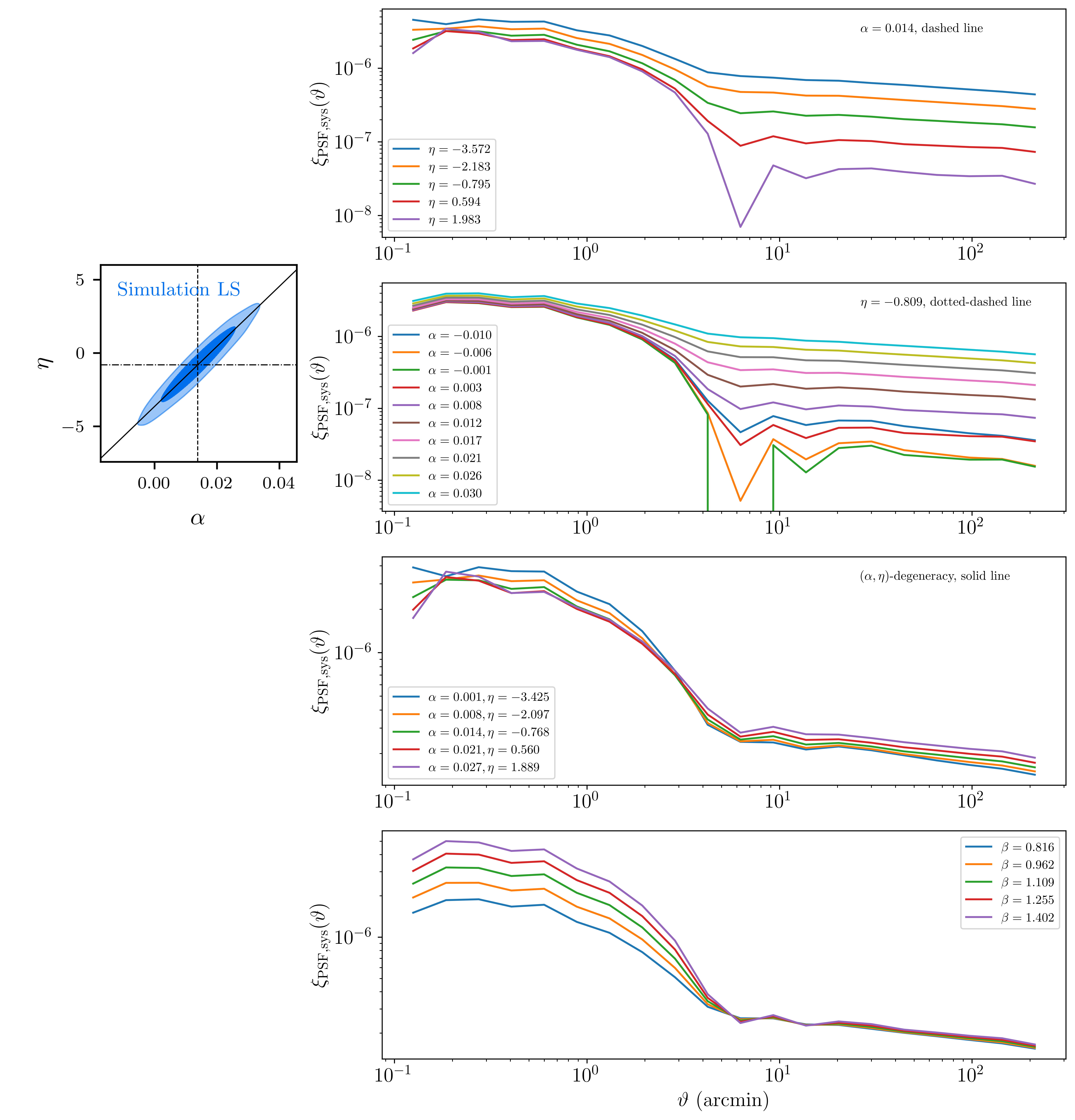}
    \caption{Dependency of the level of systematics on the parameters $\alpha$, $\beta$ and $\eta$ of the PSF error model. \textit{Upper panel:} Dependency on $\eta$ with $\alpha$ and $\beta$ fixed. It follows the dotted line on the left panel. \textit{Middle upper panel:} Dependency on $\alpha$ with $\beta$ and $\eta$ fixed. It follows the dashdotted line on the left panel. \textit{Middle lower panel:} Dependency following the degeneracy (solid line on the left panel) in the $(\alpha, \eta)$ plane with $\beta$ fixed. \textit{Lower panel:} Dependency on $\beta$ with $\alpha$ and $\eta$ fixed.}
    \label{fig:xi_sys_dependency}
\end{figure*}

In this section, we first describe the weak-lensing catalogue used in this study together with its characteristics (see Sec.~\ref{seq:wl_catalog}). We then provide insights into the simulations used to estimate the covariance for comparison with our semi-analytical method (see Sec.~\ref{seq:sim}).

\subsection{Weak-lensing catalogues}\label{seq:wl_catalog}

Our study is based on the weak-lensing shear catalogue from the Ultraviolet Near-Infrared Northern Sky Survey (UNIONS). UNIONS is an ongoing survey that targets an area of $4,800$ deg$^2$ \citep{gwynUNIONS}. UNIONS combines multi-band photometric images from multiple telescopes located in Mauna Kea. The Canada-France Hawai'i Telescope (CFHT) provides $u$- and $r$-band images \citep{2017arXiv170806356I}. This part of the survey is called the Canada-France Imaging Survey (CFIS) and is used to measure the shape of galaxies using the $r$-band. The Panoramic Survey Telescope and Rapid Response System (Pan-STARRS) provides the $i$- and $z$-band and Subaru, which takes images in the $z$-band in the framework of WISHES (Wide Imaging with Subaru HSC of the Euclid Sky), and the $g$-band Waterloo Hawai'i Ifa Survey (WHIGS). UNIONS is part of the Euclid survey and provides wide-field band observations that will contribute to obtain Euclid's photo-$z$'s in the Northern sky, together with the Euclid infrared bands. For the UNIONS v1.3 galaxy shape catalogue, the effective covered area, to which a conservative mask was applied, is around $A \sim 2,117$ deg$^2$ .

Shape measurement was performed with \textsc{ShapePipe} \citep{2022A&A...664A.141F}. A first version of the ShapePipe catalogue, presented in \cite{UNIONS_Guinot_SP}, covered $1,500$ deg$^2$. This first version of the catalogue used PSFex \citep{bertinAutomatedMorphometrySExtractor2011} to model the PSF. A more recent processing (version number v1.3) was performed containing $83,812,739$ galaxies over $3,200$ deg$^2$ of effective sky area corresponding to the available data in 2022 at the time of the processing. This more recent processing relies on MCCD \citep{MCCD21} to model the PSF. This model builds a non-parametric multi-CCD model of the PSF over the focal plane. To obtain the parameters of the PSF model, stars are selected on the individual exposures. The star sample is selected on the stellar locus in the size-magnitude diagrams. They are split into a training sample (80\%) and a validation sample (20\%). The PSF model is obtained by optimization using the training sample. The calibration of the galaxy ellipticities is performed using \textsc{Metacalibration} \citep{2017arXiv170202600H, 2017ApJ...841...24S}. This catalogue was used recently in \cite{2024arXiv240210740L} and \cite{ZK24} to measure halo masses of AGN samples. In this work we use this v1.3 shear catalogue to show the performance of our semi-analytical covariance. We also use a catalogue of $5,259,788$ validation stars.

\subsection{Simulations}\label{seq:sim}

To compute the covariance of the $\tau$-statistics using simulations, we use the software \textsc{GLASS}\footnote{Link to \href{https://GLASS.readthedocs.io/stable/}{\textsc{GLASS} documentation}} \citep{tessoreGLASSGeneratorLarge2023a}. \textsc{GLASS} provides lognormal simulations of the full sky density field and can produce galaxy catalogs sampled accordingly from this density field. \textsc{GLASS} is built to have accurate two-points statistics. The advantage of \textsc{GLASS} is that it is significantly faster than $N$-body simulations and is thus able to produce a sufficient number of simulations to adapt to the shape noise, mask or effective number of galaxies and stars of different catalogs.

We compute the effective number of galaxies/stars $n_\textrm{eff}$
and shape noise $\sigma_e$ using the following expressions from \cite{heymansCFHTLenSCanadaFrance2012}:
\begin{align}
n_{\rm eff} &= \frac{1}{A}\frac{(\sum w_i)^2}{\sum w_i^2};\\
\sigma_{e}^2 &= \frac{1}{2}\left[\frac{\sum (w_i e_{i,1})^2}{\sum w_i^2} + \frac{\sum (w_i e_{i, 2})^2}{\sum w_i^2} \right].
\end{align}
The effective number of galaxies and reserved stars are, respectively, $n_{\rm eff, gal}\sim 7.6$ arcmin$^{-2}$ and $n_{\rm eff, *} \sim 0.68$ arcmin$^{-2}$. The associated galaxy shape noise amounts to $\sigma_{e} = 0.31$. Our data vector, which is the left-hand side of Eq.~\eqref{eq:tau_matrix}, has a size of $3 \times 20 = 60$. To obtain an accurate (inverse) covariance matrix \citep{Siskind72,KIC22}, we produce $300$ simulated galaxy catalogs with the same footprint, redshift distribution and statistical properties as the data. We underline that only the galaxy catalog is simulated, and that the real star catalog is used to compute the $\rho$- and $\tau$-statistics on the simulations. As a result, the correlation between galaxies and stars is not taken into account in the covariance obtained from simulations.

\section{Results}\label{seq: results}

Here we present our results for the estimation of PSF systematics. We compare parameter constraints obtained with the semi-analytical covariance computed in Sect.~\ref{seq: analytical covariance} with results using a covariance from the data with jackknife resampling (Sect.~\ref{seq:jacknife_resampling}), and from mock simulations (see Sect.~\ref{seq:sim}).

\subsection{Comparison of the correlation matrices for different methods}

We first present a comparison of the different covariance matrices obtained using semi-analytical, jackknife resampling and simulations modelling (see Fig.~\ref{fig:corr_matrix_tau}) The three correlation matrices are in good agreement. The jackknife resampling covariance matrix displays the largest amount of noise. The simulation-based covariance seems to contain additional correlations between $\tau_2$ and $\tau_5$, and anti-correlations between $\tau_0$ and $\tau_2$ on large scales. This might be because we did not simulate stars to estimate the covariance. As the galaxy and star catalogs used to compute the galaxy-PSF correlations are independent, the terms $\xi_\pm^{\gamma b}(ik) \xi_{\pm}^{\gamma c}(jl)$ in Eq.~\eqref{eq: Covariance} are consistent with zero.

Fig.~\ref{fig:diag_cov} shows the diagonal components of the covariance matrix for each method including the contribution from the different terms appearing in Eq.~\eqref{eq:cov_abcd} and the shot noise computed with \textsc{TreeCorr}. We see some discrepancy for $\tau_0$ on large scales between the semi-analytical covariance and the one obtained with simulations. This discrepancy might be due to the interpolation used to compute the semi-analytical covariance or on assumptions made to derive the semi-analytical expression (e.g. simple geometry of the survey or Gaussianity of the fields). This might also be due to underestimation of the covariance with jackknife resampling \cite{2016MNRAS.456.2662F} and simulations, as all the terms are not taken into account.

Nevertheless, Figure \ref{fig:contours_lq_sp_v1.4} presents constraints obtained on $\alpha$, $\beta$ and $\eta$ using the least-square method and MCMC. The values of the parameters are summarized in Table \ref{tab:parameter values}. We see that the contours obtained with the three covariance matrices are in very good agreement for both least-squares and MCMC. We observe a shift towards lower values of $\beta$ when using the semi-analytical covariance matrix compared to the other two. This shift will appear in the systematic error presented in Section \ref{sec:xi_sys_level}. We also observe a small shift of the predicted values of $\alpha$ and $\eta$ using the semi-analytical covariance resulting in the shift of the contour in the $(\alpha, \eta)$-plane. This shift occurs in the direction of the degeneracy between $\alpha$ and $\eta$ and, as we will see in the next section, this will thus have no effect on the level of systematic error. The contours obtained using MCMC look similar to the one obtained from least-squares. This will be further studied in Section \ref{seq:comparison LQ MCMC}.

Figure \ref{fig:best_fit} shows the best-fit curves of the $\tau$-statistics for each method to estimate the covariance. We observe that the agreement between the three covariances is good and provides a convincing fit to the data. A slight discrepancy can be observed on large scales for $\tau_2$. The reason for this is the high correlation between the three $\tau$-statistics on large scales, which prevents the model to fit perfectly all three $\tau$-statistics at the same time.

\subsection{Comparison of the predicted level of systematics for the different covariance matrices}\label{sec:xi_sys_level}

In the previous subsection, we saw that the fitted values of $\alpha$, $\beta$ and $\eta$ are consistent between the different covariance matrices used for the $\tau-$statistics. There are however slight shifts in the contours and it is therefore important to check whether those shifts yield significantly different estimates of the level of systematic error using Eq.~\eqref{eq:xi_sys}. Figure \ref{fig:systematic error level} shows the level of systematic due to the PSF for the `+'-component of the two-point correlation function $\xi^{\gamma \gamma}_+(\vartheta)$ using both the least-squares solution and MCMC. The error bars correspond to the 68\% level of confidence. Using both methods, we see that the systematic levels are in agreement, particularly on large scales. A slight difference can be observed on small scales between the three methods. This difference can be explained by the slight shift in the value of $\beta$ between the three methods. Indeed, as can be seen in Figure \ref{fig:xi_sys_dependency}, the level of systematics on small scales is sensitive to the value of $\beta$ (see the lower panel). As a result, an offset in the estimated value of $\beta$ will result in more or less systematics on those scales. However, the slight shift in $\alpha$ and $\eta$ cannot be observed looking at the level of systematics $\xi_{\rm PSF, sys}$. From the contours in Figure \ref{fig:contours_lq_sp_v1.4}, we see that $\alpha$ and $\eta$ are positively correlated. This correlation is expected because $\alpha$ corresponds to the leakage term $\ep$ and $\eta$ to the size residual. To make the latter a spin-2 quantity it is multiplied by $e^*$ which, due to the small size of the residuals, carries also information on the leakage. The third panel of the Figure \ref{fig:xi_sys_dependency} shows that along this degeneracy in the $(\alpha, \eta)$-plane, the estimated level of systematics $\xi_{\rm PSF, sys}$ is similar. The upper panels of Figure \ref{fig:xi_sys_dependency} show that $\alpha$ and $\eta$ mostly influence the level of systematics on large scales when we don't follow their degeneracy in the $(\alpha, \eta)$-plane. A redefinition of the error model \eqref{eq:xi_sys} will be explored in Section \ref{seq:new_tau}. We will show that it can break the degeneracy between $\alpha$ and $\eta$ with no significant modification of the estimated level of systematic error $\xi_{\rm PSF, sys}$. We finally underline that the dependencies on the systematic level drawn in this section might be survey-specific as it depends on the amplitude of the $\rho$-statistics. As the choice of PSF model and shape measurement methods vary from one survey to the other, $\rho$-statistics might show different amplitudes associated with the leakage or the residuals leading to different interpretation. Nevertheless, we deem it to be important to PSF systematic and their scale-dependence, which we demonstrate here, using UNIONS data, making use of $\rho$- and $\tau$-statistics.

\subsection{Comparison of the least-squares and MCMC methods}\label{seq:comparison LQ MCMC}

\begin{figure}
    \centering
    \includegraphics[width=\linewidth]{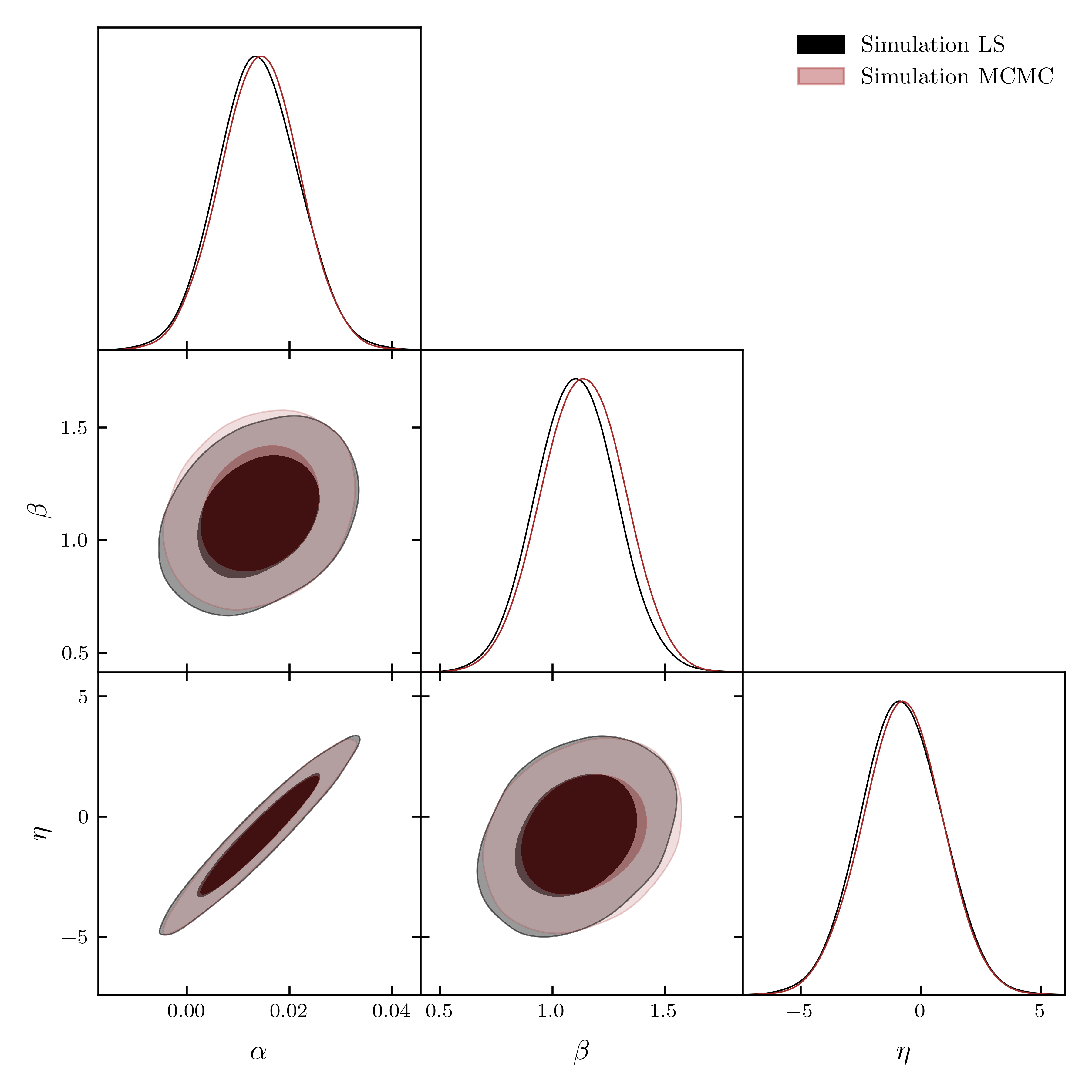}
    \caption{Constraints obtained on $\alpha$, $\beta$ and $\eta$ using least-squares (black) and MCMC (brown). The covariance used to perform the fit is obtained from \textsc{GLASS} simulations.}
    \label{fig:lq_vs_emcee}
\end{figure}

\begin{figure}
    \centering
    \includegraphics[width=\linewidth]{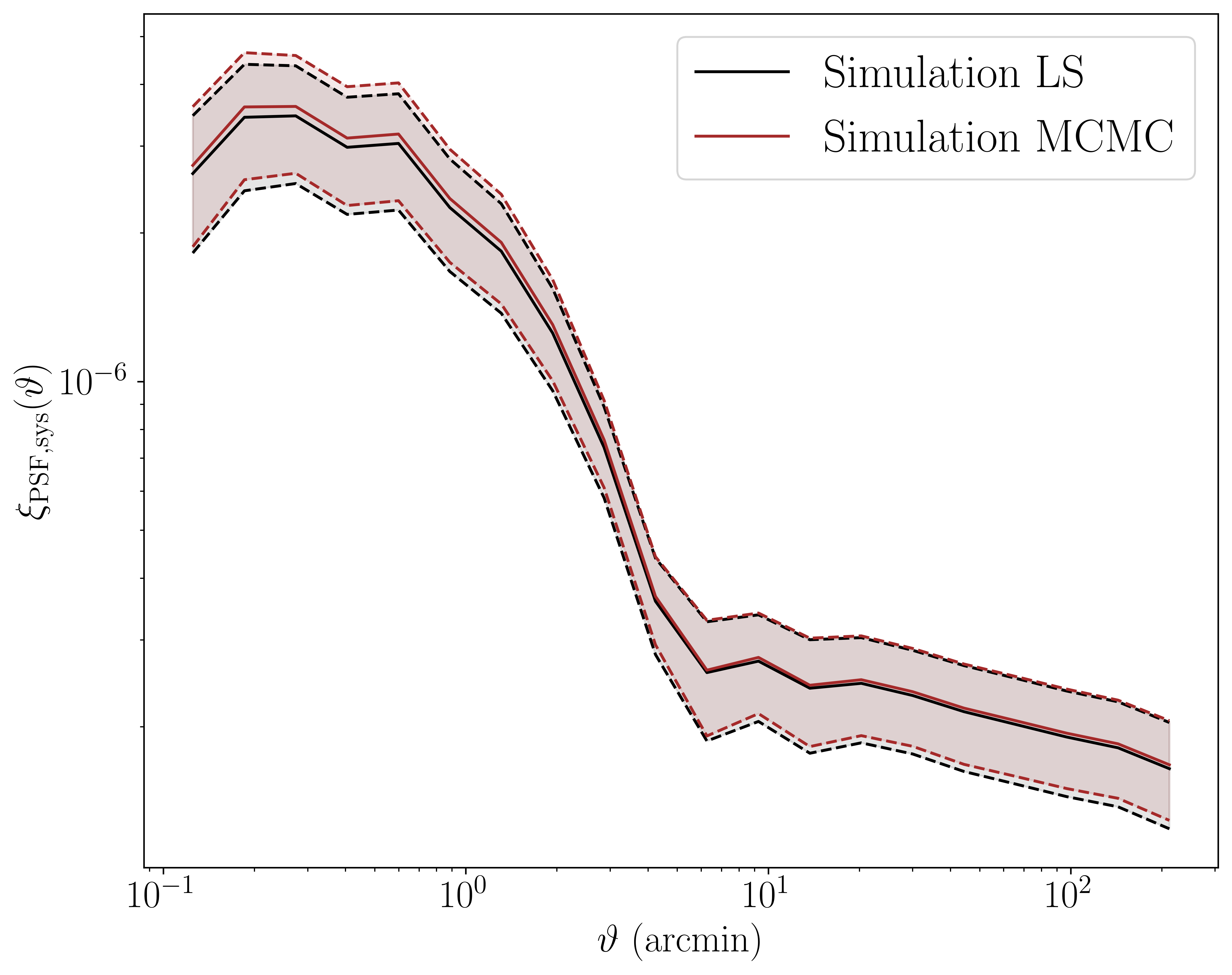}
    \caption{Level of systematics obtained using a covariance matrix computed from simulations for parameter estimation using least-squares (black) and MCMC (brown).}
    \label{fig:xi_sys_lq_vs_emcee}
\end{figure}

We now investigate differences between the contours obtained using least-squares and MCMC. Figure \ref{fig:lq_vs_emcee} shows the constraints for $\alpha$, $\beta$ and $\eta$ and Figure \ref{fig:xi_sys_lq_vs_emcee} shows the corresponding level of systematic error. The comparison is performed using the covariance matrix obtained from simulations. The contours are very similar and the corresponding level of systematic error is in very good agreement. We underline however that the least-square method relies on a frequentist approach in the sense that the parameters are sampled by solving the least-square problem sampling a $\rho$- and $\tau$-statistics realisation from their respective covariances. We stress that our work addresses the issue of covariance estimation for the $\tau$-statistics but not for the $\rho$-statistics. In our case, we used the covariance matrix obtained from jackknife resampling to sample the $\rho$-statistics and checked that the contours were little sensitive to the noise in the covariance estimate. Differences could however be observed in the size of the error bars between the method using least-squares, and therefore relying on the covariance estimate of the $\rho$-statistics, and the method using MCMC. Least-squares and MCMC give also similar results using the semi-analytical or jackknife covariance matrices. The difference in the size of the error bars between MCMC and least-squares is more pronounced when using a jackknife covariance.

\section{Discussion}\label{seq:discussion}

\subsection{A semi-analytical $\tau$-covariance for fast and accurate diagnostics}

In the previous sections, we have demonstrated that the semi-analytical covariance introduced in this work can provide a similarly precise estimate of the level of systematic error than using a covariance computed from simulations or with jackknife resampling. Despite showing a larger $\chi^2$ for the best-fit parameters than the other two methods, the semi-analytical covariance provides a similar fit to the data (See Figure \ref{fig:best_fit}) and contours consistent with the other two methods (See Figure \ref{fig:contours_lq_sp_v1.4}). The main advantage of the semi-analytical covariance compared to simulations or jackknife resampling is its reduced computation time. It does not require simulations for each catalog and therefore allows one to save a significant amount of computation time and storage. This in particularly important in a tomographic analysis where the size of the $\tau$-data vector and therefore the amount of simulations required to estimate the covariance can be very large. For example, the time required to compute the semi-analytical covariance matrix used in this paper on 48 cores of an Intel Xeon-G 5220R processor is 27 minutes compared to 227 minutes using jackknife resampling. Our analytical approach also explicitly contains cross-correlation terms between galaxy and stars that are not contained in the \textsc{GLASS} simulations. Moreover, it does not depend on the way the sky is patched contrary to jackknife resampling. It can also be used on small sky areas where jackknife resampling becomes complicated due to the small number of objects in each patch and the difficulty to find a trade-off between number of patches and number of objets per patch. It thus allows us to compute local estimates of PSF systematics and, for example, to compare the level of systematics in different patches of a survey. In a nutshell, the semi-analytical covariance for the $\tau$-statistics is a powerful tool to perform PSF diagnostics on several galaxy and star catalogs obtained with e.g.~different signal-to-noise ratio or size cuts.

\subsection{Towards a redefinition of the $\tau$-statistics}\label{seq:new_tau}

\begin{figure*}
    \centering
    \includegraphics[width=1\linewidth]{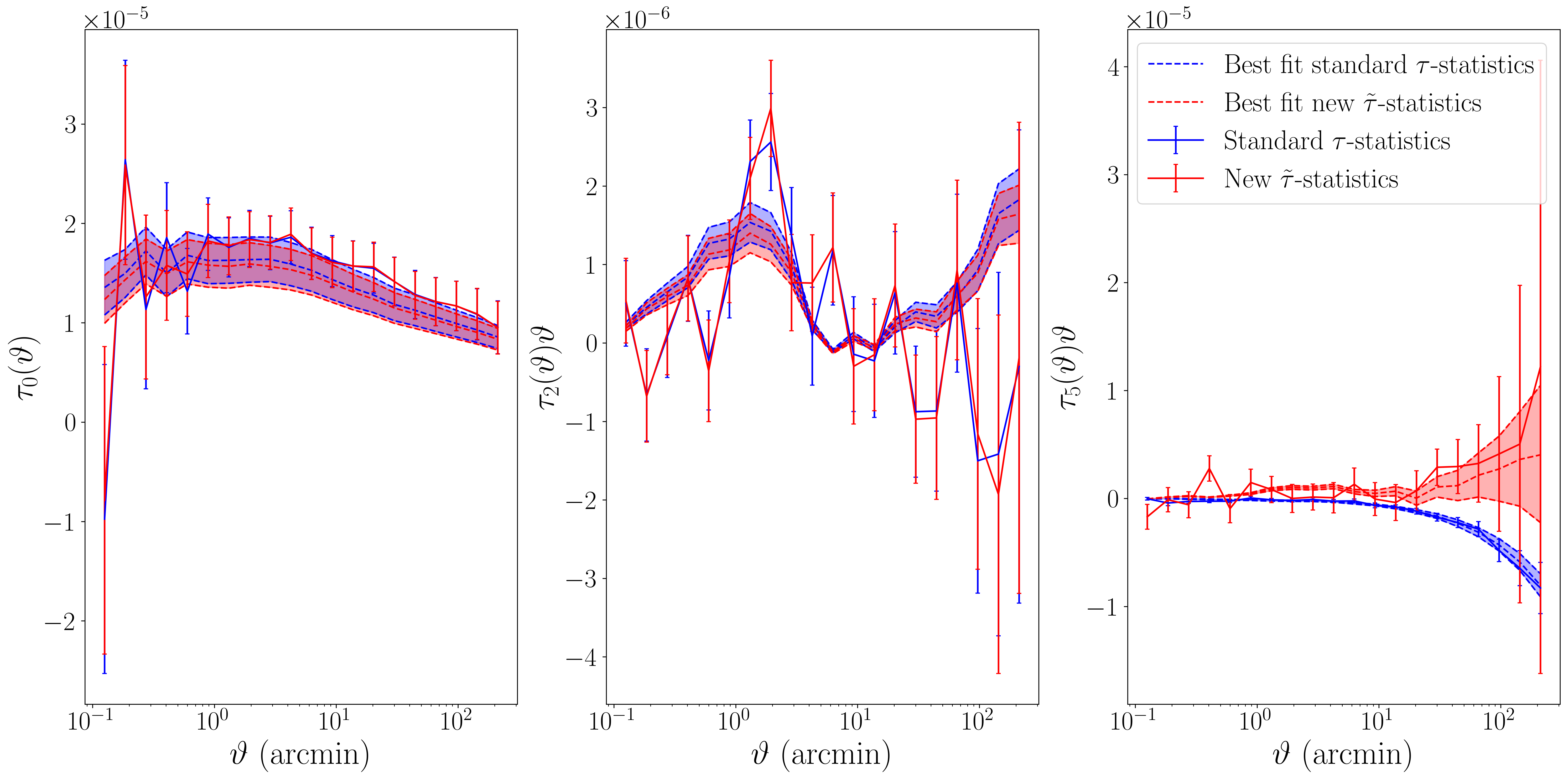}
    \caption{Best-fit model of the $\tau$- and $\tilde \tau$-statistics obtained using a least-squares sampling method. We show the 68\% confidence region.}
    \label{fig:best_fit_new_tau}
\end{figure*}

\begin{figure}
    \centering
    \includegraphics[width=\linewidth]{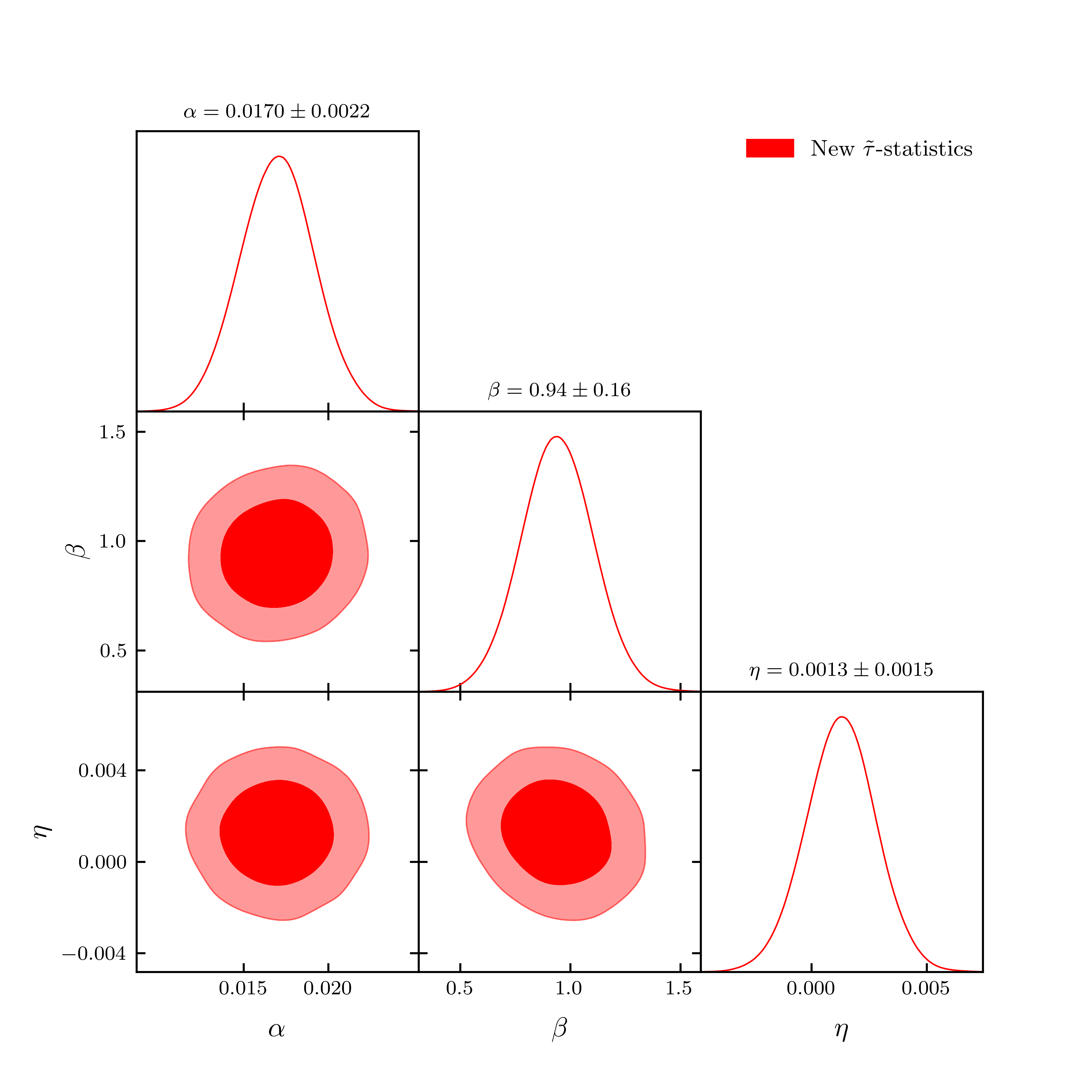}
    \caption{Constraints obtained on $\alpha$, $\beta$ and $\eta$ using least-squares using the redefined $\tilde \tau$-statistics.}
    \label{fig:contours_new_tau}
\end{figure}

\begin{figure}
    \centering
    \includegraphics[width=\linewidth]{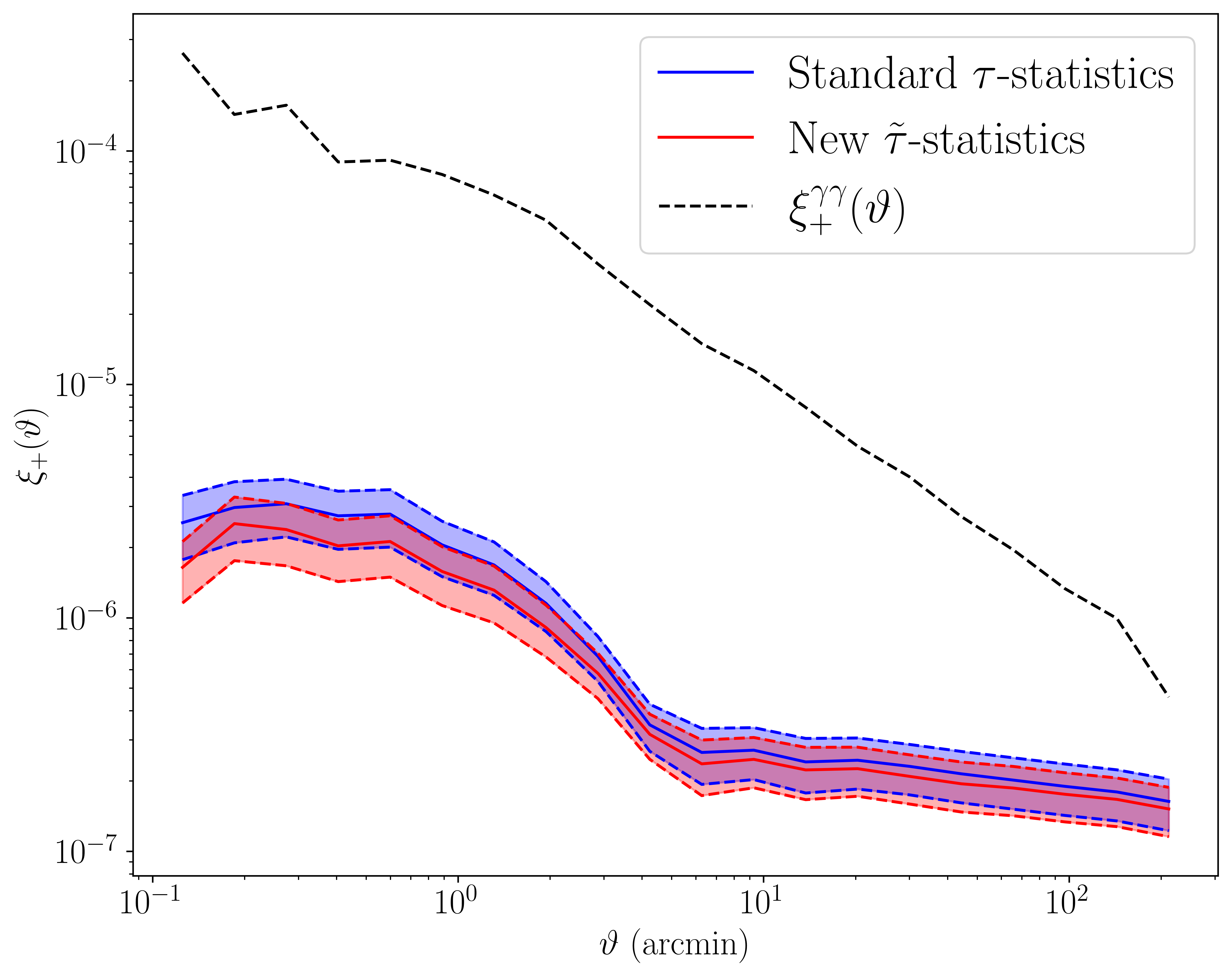}
    \caption{Level of systematics obtained using least-squares sampling method using standard $\tau$-statistics and the redefined $\tilde \tau$-statistics (see Eq.~\eqref{eq:tilde tau}).}
    \label{fig:xi_sys_new_tau}
\end{figure}

In Section \ref{sec:xi_sys_level}, we shed light on the degeneracy between the leakage parameter $\alpha$ and the size residual $\eta$ that appear with UNIONS data. This degeneracy is expected as the size residual is multiplied by the ellipticity of stars $e^*$ to yield a spin-2 field. This allows spin consistency between all terms in Eq.~\eqref{eq:e_sys} although the introduction of this parameter does not provide additional information \citep[see e.g.][]{giblinKiDS1000CatalogueWeak2021a, zhangGeneralFrameworkRemoving2023, 10.1093/mnras/stab918}. In this section, we redefine the $\rho$- and $\tau$-statistics by replacing the $e^*$ factor in front of the size residual with a new complex spin-2 field $f$ such that:
\begin{align}
    \tilde e^\textrm{PSF, sys} = \alpha \ep + \beta \, \delta \ep + \eta \, \delta \tilde T^{\rm p}, 
    \label{eq:e_sys_tilde}
\end{align}
where $\delta \tilde T^{\rm p} = f (T^* -\Tp)/T^*$.
The field $f$ is implicitly defined, with respect to the pairs, such that, when computing the $\rho$- and $\tau$-correlations, only the tangential direction contributes at star positions and for a given polar angle $\phi$. $f$ thus reverts to the unit vector in the tangential direction, $f = \hat e_\textrm{t}$. Thus the correlation function between a spin-2 field $a$ and the product of $f$ with a scalar field $s$ is obtained as follows:
\begin{align}
    \xi^{a [fs]}_\pm(\vartheta) &= \langle a_t f_t s \rangle(\vartheta) \pm \underbrace{\langle a_\times f_\times s \rangle (\vartheta)}_{=0} \nonumber \\
    &= \langle a_t s \rangle(\vartheta).
\end{align}
The `$+$' and `$-$' components of the correlation function are thus degenerate and correspond to the correlation between the tangential component of the spin-2 field $a$ at each objects position with respect to the scalar field $s$ which corresponds to what we expect. The $\rho$- and $\tau$-statistics involving the size residuals are redefined as well with:
\begin{align}
    &\tilde \rho_3(\vartheta) = \langle \delta \tilde T^{\rm p} \delta \tilde T^{\rm p} \rangle; \quad
    \tilde \rho_4(\vartheta) = \langle \delta \ep \delta \tilde T^{\rm p} \rangle; \quad
    \tilde \rho_5(\vartheta) = \langle \ep \delta \tilde T^{\rm p} \rangle,
\end{align}
and
\begin{equation}\label{eq:tilde tau}
    \tilde \tau_5(\vartheta) = \langle e \delta \tilde T^{\rm p} \rangle.
\end{equation}
With that, the two-point correlations $\tilde \rho_4, \tilde \rho_5$ and $\tilde \tau_5$ correspond to the expected tangential ellipticity around stars weighted by $\delta T$, and $\tilde \rho_3$ is the $\delta T$ scalar auto-correlation function.
In practice, we use the \textsc{TreeCorr} \textsc{KKCorrelation} class for $\tilde \rho_3$ and \textsc{GKCorrelation} for $\tilde \rho_4, \tilde \rho_5$ and $\tilde \tau_5$ and set $\delta T$ as the scalar field `K' for the correlation.
With this new definition, $\tilde \tau_5$ provides a null-test related to the quality of the size estimation of the PSF whereas $\tau_5$ carried information on both the size residual and the leakage due to the $e^*$ factor.
Figures \ref{fig:best_fit_new_tau} and \ref{fig:contours_new_tau} show that the $\tilde \tau$-statistics provide an accurate fit to the data and remove the degeneracy between the leakage parameter $\alpha$ and the size residuals parameter $\eta$. $\tilde \tau_5$ is the only modified $\tau$-statistics and provides a null-test to check if the size residuals contribute significantly to an additive bias. With our data, it seems to show that the leakage is the main contributor to correlation between galaxies and stars shape whereas the correlation with the size residuals is consistent with zero (see right panel in Fig.~\ref{fig:best_fit_new_tau}). The covariance matrix of the $\tilde \tau$-statistics is estimated using jackknife resampling. Figure \ref{fig:xi_sys_new_tau} shows that the level of systematics obtained with the $\tilde \tau$-statistics matches the one obtained from the standard $\tau$-statistics at the 68\% confidence level. With the $\tilde \tau$-statistics we are thus able to break the degeneracy between $\alpha$ and $\eta$ by disentangling the information carried by $\tau_0$ and $\tau_5$. Hence, we advocate that $\tilde \tau$-statistics provide a more interpretable description of PSF systematics, in particular when the size residuals are not very small, and should be preferred over standard $\tau$-statistics when a strong degeneracy between $\alpha$ and $\eta$ is observed. The semi-analytical covariance matrix introduced in Sect.~\ref{seq: analytical covariance} should be revised to compute the auto- and cross-correlations involving $\tilde \tau_5$. We leave this for future work.

\section{Conclusions}
\label{seq: Conclusions}

The goal of this paper is to provide a methodology to compute fast and accurate estimates of PSF systematics that pollute two-point cosmic shear correlation function. To that end, we used the so-called $\rho$- and $\tau$-statistics, and developed a semi-analytical covariance of the $\tau$-statistics to obtain fast and accurate estimates of PSF systematics for a given catalog of galaxies (Sect.~\ref{seq: analytical covariance}).

In addition, we estimated the PSF parameters as the exact solution of a least-square problem, and explore the credible region of parameter space via sampling of the covariance matrices of the $\tau$- and $\rho$-statistics.

We applied our methodology to UNIONS data, where the PSF was obtained with the data-driven PSF model MCCD (Section \ref{seq: data and simulations}). We find a good agreement of our semi-analytical covariance with estimates using simulations and jackknife resampling (Sect.~\ref{seq: results}). We stress the comparatively low computation time to calculate the semi-analytical covariance matrix, compared to jackknife or simulation-based estimates with acceptable low sampling noise. The speed-up is of the order 8 between semi-analytical and jackknife covariance.
We note that the semi-analytical covariance contains galaxy-PSF cross-correlation terms that are missing from state-of-the-art simulations. 

The inferred PSF systematics parameters are in excellent agreement between all three covariance matrix inputs, even though with the semi-analytical covariance the reduced $\chi^2$ value of the best-fit model is the highest of the three. 

We find excellent agreement of our least-square parameter estimation and exploration with Monte-Carlo Markov chain sampling, at an order of 3 times lower computation time. The total gain in computing time using least-squares with the semi-analytical covariance over MCMC sampling using jackknife or simulations is a factor 8.2. 

We also observed that our analysis finds an important degeneracy between the parameter $\alpha$, measuring the amplitude of the ``PSF leakage'', and the parameter $\eta$ associated to the size residuals of the PSF (Section \ref{seq:discussion}). This degeneracy arises naturally as the size residual is multiplied by the ellipticity of stars to yield a spin-2 field. We explored a redefinition of the $\tau$-statistics that considers the size residuals as a scalar field to disentangle the PSF systematics coming from PSF leakage and from the size residuals. We argue that, even though the error model loses physical consistency, it allows to recover results similar to the standard $\tau$-statistics without the degeneracy in the $(\alpha, \eta)$-plane. In cases where this degeneracy appears, one should prefer to use this redefinition to know whether leakage or size residuals dominate the PSF systematics. We also highlight that we only considered PSF second moments whereas previous work have shown that fourth moments can carry significant information on PSF systematics that is not taken into account in our context \citep[See e.g.][for more information]{zhangGeneralFrameworkRemoving2023}. We underline that the mathematical framework to compute a semi-analytical covariance matrix with $\tau$-statistics extended to fourth moment is analogous to the one introduced in this work and can be easily transposed if needed.

This work will provide useful tools to perform the analysis of the ongoing UNIONS survey but also for future surveys (e.g. Euclid or LSST). This can be a useful tool in the era of Stage IV large-scale structure survey to perform fast PSF diagnostics on catalogs obtained either on different patches of the sky or with different cuts that could influence PSF systematics. It will be crucial to understand those systematics correctly as the wide area coverage and increased depth will likely yield statistical errors that are negligible compared to the systematics for the shear-shear two-point correlation function. In the context of the Euclid mission, it will also be a useful tool to validate the shape measurement algorithm and the PSF model. Finally, as soon as the analysis pipeline is frozen, it provides useful estimate of the PSF systematics that can be taken into account in the modelling when performing Markov Chain Monte Carlo estimations of the cosmological parameters or calibrated for beforehand \citep{liKiDSLegacyCalibrationUnifying2023}. This methodology will also be useful in a context where forward models become essential tool to validate analysis pipelines e.g. for 3x2 point statistics \citep{amonConsistentLensingClustering2023} or to perform inference using Simulation-Based methods \citep{vonwietersheim-kramstaKiDSSBISimulationBasedInference2024, jeffreyDarkEnergySurvey2024, gattiDarkEnergySurvey2023} where all systematics must be forward-modeled correctly to avoid model misspecification and guarantee unbiased inference \citep{cannonInvestigatingImpactModel2022a}.

For the sake of reproducible research, the figures shown in this work can be reproduced using the code available on the GitHub repository associated with the paper \href{https://github.com/sachaguer/tau\_stats\_paper}{\faGithub}. The repository contains the covariance matrices computed using simulations, jackknife resampling, and the analytical expressions, as well as the samples obtained from these matrices. Companion notebooks allow the user to create the figures from the provided data. The code used to compute the semi-analytical covariance matrix is publicly available on GitHub\href{https://github.com/martinkilbinger/shear\_psf\_leakage}{\faGithub}.

\begin{acknowledgements}
We are honored and grateful for the opportunity of observing
the Universe from Maunakea and Haleakala, which both have cultural, historical
and natural significance in Hawaii. This work is based on data obtained as part of
the Canada-France Imaging Survey, a CFHT large program of the National Research Council of Canada and the French Centre National de la Recherche Scientifique. Based on observations obtained with MegaPrime/MegaCam, a joint
project of CFHT and CEA Saclay, at the Canada-France-Hawaii Telescope
(CFHT) which is operated by the National Research Council (NRC) of Canada,
the Institut National des Science de l’Univers (INSU) of the Centre National de
la Recherche Scientifique (CNRS) of France, and the University of Hawaii. This
research used the facilities of the Canadian Astronomy Data Centre operated
by the National Research Council of Canada with the support of the Canadian
Space Agency. This research is based in part on data collected at Subaru Telescope, which is operated by the National Astronomical Observatory of Japan.
Pan-STARRS is a project of the Institute for Astronomy of the University of
Hawaii, and is supported by the NASA SSO Near Earth Observation Program under grants 80NSSC18K0971, NNX14AM74G, NNX12AR65G, NNX13AQ47G,
NNX08AR22G, 80NSSC21K1572 and by the State of Hawaii. LB is supported
by the PRIN 2022 project EMC2 - Euclid Mission Cluster Cosmology: unlock
the full cosmological utility of the Euclid photometric cluster catalog (code no.
J53D23001620006). This work was made possible by utilising the CANDIDE cluster at the Institut d’Astrophysique de Paris. The cluster was funded through grants from the PNCG, CNES, DIM-ACAV, the Euclid Consortium, and the Danish National Research Foundation Cosmic Dawn Center (DNRF140). It is maintained by Stephane Rouberol. We also gratefully acknowledge support from the CNRS/IN2P3 Computing Center (Lyon - France) for providing computing and data-processing resources needed for this work.
H. Hildebrandt is supported by a DFG Heisenberg grant (Hi 1495/5-1), the DFG Collaborative Research Center SFB1491, an ERC Consolidator Grant (No. 770935), and the DLR project 50QE2305
\end{acknowledgements}

\bibliographystyle{aa}
\bibliography{astro, tau_stats_paper}

\begin{appendix}

\section{Derivation of the two-point spin-2 correlators}\label{sec:terms_der}

Here we detail the derivation of the relations Eq.~\eqref{eq:terms_ab}.
We first define the parity-violating correlation function $\xi_\times$, which
mixes tangential and cross components. Another parity-variant function exists, which we call $\xi_\ast$. Their expressions are
\begin{align}\label{eq: xi_star_cross}
    \xi_\ast^{ab}(\vartheta)
    &
    = - \left\langle a_\textrm{t} b_\times \right\rangle(\vartheta)
        + \left\langle a_\times b_\textrm{t} \right\rangle(\vartheta)
    \nonumber \\
    &=
    - \left\langle a_1 b_2 \right\rangle(\vartheta) + \left\langle a_2 b_1 \right\rangle(\vartheta)
    \nonumber\\
    &=
    \Im \left[
    \left\langle a b^\ast \right\rangle(\vartheta)
    \right] ;
    \nonumber \\
    \xi_\times^{ab}(\vartheta)
    & = 
    \left\langle a_\textrm{t} b_\times \right\rangle(\vartheta) 
    + \left\langle a_\times b_\textrm{t} \right\rangle(\vartheta)
    \nonumber \\
    &=
    \left\langle
    \left[
    - \left(a_1 b_1 \right)(\vec \vartheta)
    + \left( a_2 b_2 \right)(\vec \vartheta)
    \right] \sin 4 \phi
    \right\rangle
    \nonumber\\
    & \quad +
    \left\langle
    \left[
    \left(a_1 b_2 \right)(\vec \vartheta)
    + \left( a_2 b_1 \right)(\vec \vartheta)
    \right] \cos 4 \phi
    \right\rangle
    \nonumber \\
    &= \Im \left\langle
        \left( ab \right) (\vec \vartheta) e^{-4 \textrm{i} \phi}
    \right\rangle
    .
\end{align}
The four real-valued correlation functions defined in Eqs.~\eqref{eq: xi_plus_minus} and \eqref{eq: xi_star_cross} can be combined into two complex correlation 
functions between the field $a_\textrm{t} + \textrm{i} a_\times = a \exp(-2 \textrm{i} \phi)$ 
and the field $b_\textrm{t} + \textrm{i} b_\times$ and its complex conjugate $b_\textrm{t} - 
\textrm{i} b_\times$, respectively; these fields are well-defined given a direction with polar angle $\phi$. The two complex correlation functions $\left\langle a b^\ast \right\rangle(\vartheta)$ and
$\left\langle (a b)(\vec \vartheta) \exp(-4 \textrm{i} \phi) \right\rangle$ were introduced in \cite{JBJ04} as the \emph{natural components} of the shear two-point correlation function. Combining Eq.~\eqref{eq:a_tx} into a complex quantity $\tilde a = a_\textrm{t} + \textrm{i} a_\times$, we can write these natural component as
\begin{align}
    \xi_+^{ab}(\vartheta) + \textrm{i} \, \xi_\ast^{ab}(\vartheta)
        & = \left\langle a b^\ast \right\rangle(\vartheta)
        = \left\langle \tilde a \tilde b^\ast \right\rangle(\vartheta) ;
        \nonumber \\
    \xi_-^{ab}(\vartheta) + \textrm{i} \, \xi_\times^{ab}(\vartheta)
        & = \left\langle a b \textrm{e}^{-4 \textrm{i} \phi} \right\rangle(\vartheta)
        = \left\langle \tilde a \tilde b \right\rangle(\vartheta) .
\end{align}

Getting back to Eqs.~\eqref{eq:terms_ab},
for the first two equal-component equations, we expand
\begin{align}
    \left\langle a_{i \alpha} b_{j \alpha} \right\rangle
    = \frac 1 2
        \left(
        \left\langle a_{i \alpha} b_{j \alpha} \right\rangle
        + \left\langle a_{i \beta} b_{j \beta} \right\rangle
        \right)
        + \frac 1 2
        \left[
        \left\langle a_{i \alpha} b_{j \alpha} \right\rangle
        - \left\langle a_{i \beta} b_{j \beta} \right\rangle
        \right] .
\end{align}
With $\alpha \neq \beta$ sum in the first round bracket
is $\xi_+^{ab}(\vartheta_{ij})$. For the difference in the second bracket we note that
when forming the combination $\xi_-^{ab}(\vartheta_{ij}) \cos 4 \phi_{ij} - \xi_\times(\vartheta_{ij}) \sin 4 \phi_{ij}$ all mixed-component terms vanish, leaving the difference of the products of equal components. The difference between $\alpha=1$ (and thus $\beta=2$) and $\alpha =2$ is a minus sign, and we recover the first two equations of Eq.~\eqref{eq:terms_ab}.

For the third and fourth, mixed-component equations, we expand in a similar way
\begin{align}
    \left\langle a_{i \alpha} b_{j \beta} \right\rangle
    = \frac 1 2
        \left(
        \left\langle a_{i \alpha} b_{j \beta} \right\rangle
        + \left\langle a_{i \beta} b_{j \alpha} \right\rangle
        \right)
        + \frac 1 2
        \left(
        \left\langle a_{i \alpha} b_{j \beta} \right\rangle
        - \left\langle a_{i \beta} b_{j \alpha} \right\rangle
        \right) .
\end{align}
The second bracket is $\xi_\ast(\vartheta_{ij})$ if $\alpha=1$, and with an additional minus sign for $\alpha=2$. The first term is symmetrical in $\alpha$ and $\beta$, and equal to
$\xi_-^{ab}(\vartheta_{ij}) \sin 4 \phi_{ij} + \xi_\times^{ab}(\vartheta_{ij}) \cos 4 \phi_{ij}$.



%
%
%
%
%
%
\section{Gaussianity test of the $\tau$-statistics}\label{seq:gauss_test}
\begin{figure*}
    \centering
    \includegraphics[width=\linewidth]{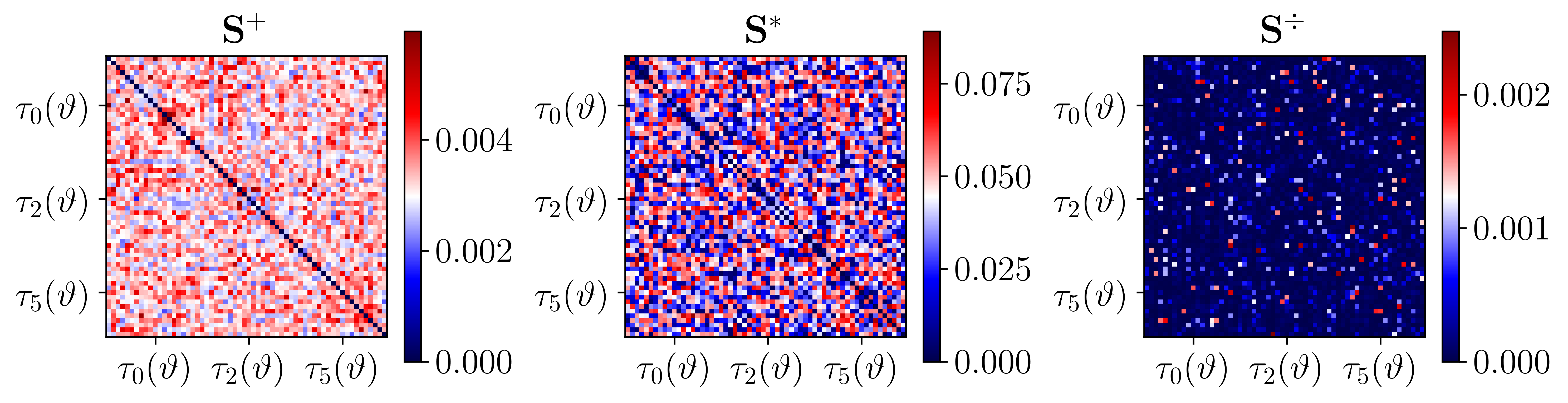}
    \caption{Transcovariance matrices $\mat S^+$, $\mat S^*$ and $\mat S^\div$ obtained from the $\tau$-statistics samples from \textsc{GLASS} simulations. Each matrix coefficient $(e,f)$ measures the amount of non-Gaussianity in the correlation between the $e$-th and the $f$-th entry of the data vector.}
    \label{fig:transcovariances}
\end{figure*}
\begin{figure*}
    \centering
    \includegraphics[width=\linewidth]{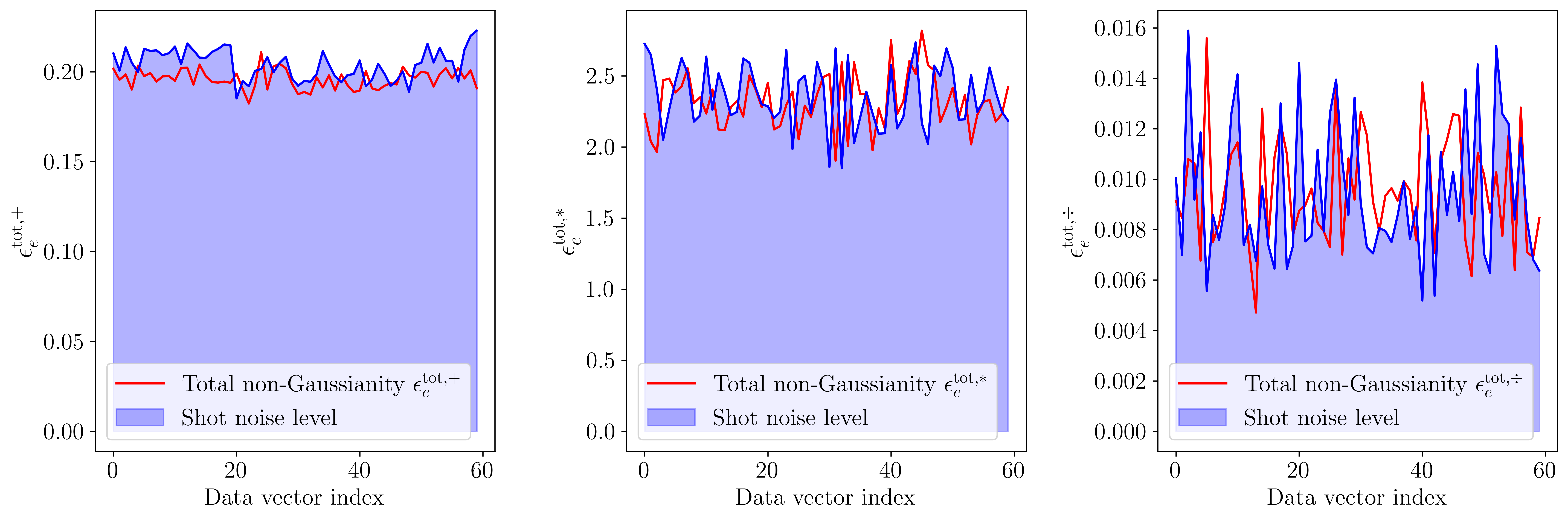}
    \caption{Total non-Gaussianity contamination at each entry of the data vector for each test. The blue region represents the shot noise due to the finite number of samples $N$ and the limited amount of bins $B$. The red lines show the non-Gaussianity obtained for the $\tau$-statistics that are not significantly above the shot noise.}
    \label{fig:total_non_gaussianity}
\end{figure*}
One of the main hypothesis underlying the expression of the semi-analytical covariance matrix in Eq.~\eqref{eq: Covariance} is that the fields considered to compute the $\tau$-statistics are Gaussian. It allows us to apply Wick's theorem in Eq.~\eqref{eq:wick} to expand the four-point correlators in a sum of two-point correlators that we are able to measure easily. To test this hypothesis, we apply the methodology of \cite{sellentinInsufficiencyArbitrarilyPrecise2018a} with similar notations. That work introduced three ``transcovariance'' matrices to detect non-Gaussianities in the likelihood. These matrices help to examine whether the sum, the ratio, and the product of two independently drawn samples of the data are distributed differently from a Gaussian distribution, a Cauchy distribution, and two linearly superposed $\chi^2_1$-random variables,\ respectively. Those distributions are expected for samples drawn from independent unit Gaussians. As we simulated $300$ independent realisations of the $\tau$-statistics using \textsc{GLASS} (see Sect.~\ref{seq:sim}), we can use them to compute the ``transcovariance'' matrices of the $\tau$-statistics.

We note with $x_i^e$ the $i^{\textrm{th}}$ realisation of element $e$ of the $\tau$-statistics data vector ($= e^\textrm{th}$ row in Eq.~\eqref{eq:tau_matrix}).
Our goal is to test whether two $\tau$-values $x^e$ and $x^f$ pass the Gaussianity tests. To that end, we first remove the Gaussian correlation between both variables with a mean-subtraction and whitening step. The whitening step is applied to the two-dimensional covariance matrix between $x^e$ and $x^f$, $C_{\alpha\beta} = \left\langle x^\alpha x^\beta \right\rangle$, for $\alpha, \beta = 1, 2$, with $x^1 = x^e$, $x^2 = x^f$.
If the data was indeed Gaussian, the sum, ratio and product of the samples should have the distributions as stated above. To test if the distributions match, we use the $N=300$ simulated and whitened $\tau$-statistics to estimate the sum, ratio and product for each pair $(e,f)$ of data vector entry. We then bin their distribution into $B$ histogram values $\mathcal{H}_b, b=1 \ldots B$. The distance from the histogram to the expected distribution is computed using the so-called Mean Integrated Squared Errors (MISE). As an example, for the sum,
\begin{align}
    s_i^{e,f} = x_i^e+x_i^f,
\end{align}
the target distribution is a Gaussian with variance $2$. It allows to define the ``transcovariance'' for the sum:
\begin{align}
    \mat S^+_{e,f} = \frac{1}{B}\sum\limits_{b=1}^{B}[\mathcal{H}_b(s_i^{e,f})-\mathcal{G}(0,2)]^2.
\end{align}
In the limit where the number of samples $N$ and the number of bins $B$ go to infinity, the weak law of great numbers ensures the convergence of the histogram to $\mathcal{G}(0,2)$ if the data only contains Gaussian correlations. In that case, $\mat S^+$ should be null. One can define in the same way transcovariance matrices $\mat S^*$ and $\mat S^\div$ for the product and the ratio of the samples. In the case of the product, the probability distribution function has no closed form as the two $\chi^2_1$ linearly superposed are not independent. It is however easy to sample from this distribution and therefore we will compute the MISE for $\mat S^*$ using a histogram sampled from the correct distribution. Fig.~\ref{fig:transcovariances} shows those transcovariance matrices obtained using \textsc{GLASS} simulations. Each matrix coefficient $(e,f)$ corresponds to the amount of non-Gaussianity in the correlation between $x^e$ and $x^f$. Because we only have access to a finite amount of samples $N$ and number of bins $B$, the estimate of the total non-Gaussian contamination $\epsilon_e^{\rm tot,+}$ is polluted by a shot noise. To characterize the noise, we draw $N$ Gaussian distributed samples with the same covariance as our dataset. From those $N$ Gaussian distributed samples, we compute calibration matrices $\mat S_{\rm cal}^+$, $\mat S_{\rm cal}^*$ and $\mat S_{\rm cal}^\div$. In particular, this calibration set allows us to choose the number of bins to compute the histogram that minimizes the amount of structure observed in the transcovariances $\mat S^*$ and $\mat S^\div$ as their distribution is peaked and can thus introduce spurious structures in the Gaussian samples transcovariances.

We can therefore use those matrices to define the total non-Gaussian contamination of the $e$-th data point as the sum over the column of the transcovariance matrix, e.g.~for the sum:
\begin{align}\label{eq:total_non_gaussian}
    \epsilon_e^{\rm tot,+} = \sum\limits_{f \ne e}\mat S^+_{e,f}.
\end{align}
One can then define a threshold, for the total non-Gaussian contamination defined in Eq.~\eqref{eq:total_non_gaussian}, beyond which the distribution of the data point is considered non-Gaussian. %
To characterize if this value is significant and points at non-Gaussian correlation, it has to be compared with the shot noise due to the limited amount of samples $N$ and bins $B$ computed using data sampled from a multivariate Gaussian distribution as described above. Fig.~\ref{fig:total_non_gaussianity} shows the total non-Gaussian contamination for each data vector entry against the shot noise computed on the Gaussian samples. We note that the contamination obtained is noisier when using the product or ratio transcovariance matrices as the Cauchy and $\chi^2_1$ are more peaked. We see that the total non-Gaussian contamination is not significantly higher than the shot noise and we thus do not detect deviations from Gaussianity in the distribution of the data. The Gaussian approximation used in the analysis in Sect.~\ref{seq:tau_stats} is thus justified.

\end{appendix}

\end{document}